\documentclass[a4paper,11pt]{article}
 
\usepackage{amsmath,amssymb,epsfig,}

\title{Link-Space and Network Analysis}
\author{David M.D. Smith, Chiu Fan Lee, Neil F. Johnson and Jukka-Pekka Onnela}
\begin{document}
\date{\today}
\maketitle
%\begin{romanpages}
%\listoffigures
%\end{romanpages}
\begin{abstract}
% Text of abstract
Many networks contain correlations and often conventional analysis is incapable of incorporating this often essential feature. In arXiv:0708.2176, we introduced the link-space formalism for analysing degree-degree correlations in evolving networks. In this extended version, we provide additional mathematical details and supplementary material. 

We explore some of the common oversights when these correlations are not taken into account, highlighting the importance of the formalism. 
The formalism is based on a statistical description of the \emph{fraction of links} $l_{i,j}$ connecting nodes of degrees $i$ and $j$. To demonstrate its use, we apply the framework to some pedagogical network models, namely, random-attachment, Barab\'asi-Albert preferential attachment and the classical Erd\H{o}s and R\'enyi random graph. For these three models the link-space matrix can be solved analytically. We apply the formalism to a simple one-parameter growing network model whose numerical solution exemplifies the effect of degree-degree correlations for the resulting degree distribution. We also employ the formalism to derive the degree distributions of two very simple network decay models, more specifically, that of random link deletion and random node deletion. The formalism allows detailed analysis of the correlations within networks and we also employ it to derive the form of a perfectly non-assortative network for arbitrary degree distribution.  
\end{abstract}
\newpage
\tableofcontents

\newpage
\section{Introduction}\label{sec:LSintro}
Networks -- in particular large networks with many nodes and links -- are attracting widespread attention. The classic reviews \cite{ABReview, Dorogovtsev, NewmanReview} with their primary focus on structural properties have been followed up by more recent ones addressing the role of dynamics, such as spreading and synchronisation processes on networks, as well as the role of weights and mesoscopic structures, i.e. cliques (fully connected subgraphs) and communities (groups of densely interconnected nodes), within networks \cite{Boccalettia,Newmanbook}. 

Although several different measures for characterising networks have been presented, for example in a recent survey \cite{Costa}, the simple concept of vertex degree remains unrivalled in its ability to capture fundamental network properties. When comparing the degrees of connected vertices, however, one often finds that they are correlated, a quality that gives rise to a rich set of phenomena \cite{Callaway2, Krapivsky, Newman}. Degree correlations constitute a central role in network characterisation and modelling but, in addition to being important in their own right, also have substantial consequences for dynamical processes unfolding on networks. Given the increasing current interest in network dynamics, understanding structural correlations remains important and timely.

In this paper we provide a detailed mathematical formalism for modelling networks with correlations. It is built around a statistical description of inter-node linkages as opposed to single-node degrees. Most works devoted to analytical calculations of correlations in models have been performed only for particular cases~\cite{Barrat1}.  We provide a brief overview of degree correlations for network structure and dynamics in Section \ref{sec:correlations} and discuss how the work presented relates to existing studies. Section~\ref{sec:notation} contains the notation used throughout the rest of the paper. We highlight the importance of acknowledging these correlations in Section~\ref{sec:Saramaki} through reviewing a model of network growth proposed by Saram\"aki and Kaski~\cite{Saramaki} (and subsequently by Evans and Saram\"aki~\cite{Evans}) which employs random walkers to grow a network. The \emph{link-space} formalism, which lies at the core of the paper, is introduced in Section \ref{sec:linkspace}. The formalism comprises a master equation description of the evolution of a very specific matrix construction which we term the \emph{link-space matrix}. The formalism can be implemented in a number of ways. To demonstrate its use, we apply it to two well-known, non-equilibrium examples, namely random-attachment and Barab\'asi-Albert (BA) preferential-attachment networks \cite{Barabasi,BarabasiDeriv2} in Section~\ref{sec:linkspace} and solve the so-called link-space matrix analytically for these models. In Section~\ref{sec:linkspace}, we also apply it to the classical equilibrium random graph of Erd\H{o}s and R\'enyi ~\cite{ER} (ER) using some of the ideas discussed in \cite{AcceleratingNetworks} which, interestingly, requires a full, time-dependent solution of the link-space master equations. The cumulative link-space introduced in Section~\ref{sec:linkspace} aids comparison between simulated and analytic link-space matrices and could, in principle, be applied to empirical networks to better ascertain their degree correlations. Analytic derivation of link-space matrices allows detailed analysis of the degree-degree correlations present within these networks as is demonstrated in Section \ref{sec:degrees}. Within Section~\ref{sec:degrees}, we also show how, for arbitrary degree distribution, a link-space matrix can be derived which has no correlations present, i.e. the link-space is representative of a perfectly non-assortative network with that degree distribution. We then consider the counter-intuitive prospect of finding steady states of decaying networks in Section \ref{sec:decay}. In Section~\ref{sec:Redner}, we discuss the Growing Network with Redirection (GNR) model of Krapivsky and Redner~\cite{Krapivsky}. In Section~\ref{sec:model}, we introduce a simple one-parameter network growth algorithm with a similar redirection process which is able to produce a wide range of degree distributions. The model is interesting in its own right in the sense that it makes use of only local information about node degrees. Here the link-space formalism allows us to identify the transition point at higher exponent degree distibutions switch to lower exponent distributions with respect to the BA model. We conclude in Section~\ref{sec:lsconclusion}. 

\section{Overview of degree correlations}\label{sec:correlations}
\begin{figure}
\begin{center}
\includegraphics[width=0.55\textwidth]{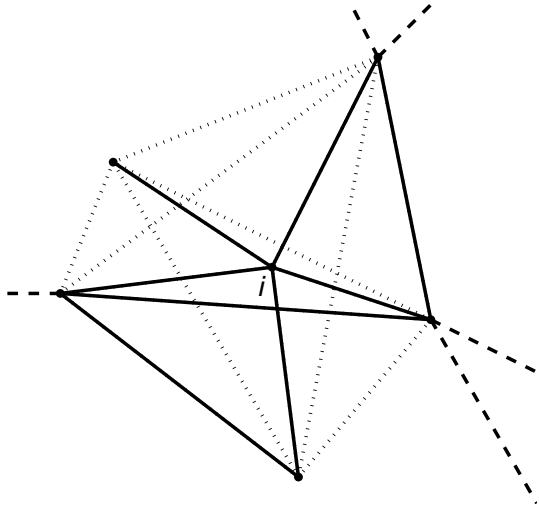}
\caption{ \label{fig:clustering1} Node $i$ has $5$ neighbours which could potentially share 10 links between them. Only three out of these $10$ possible links exist between the neighbours of $i$. Its clustering coefficient is then expressed $C_i~=~0.3$.}
\end{center} 
\end{figure}

A discussion of correlations might naturally start with the phenomenon of \emph{clustering}. The highly influential paper of Barab\'asi and Albert and many subsequent papers focussed on the modelling and analysis of models which replicate  power-law degree distributions observed in many real-world systems \cite{Barabasi, DorogovtsevDeriv, NewmanReview}. However, the BA preferential attachment mechanism lacks certain key features observed in many of those systems, one such feature being \emph{clustering}. Clustering is an important local statistic in many networks and reflects the connectedness of neighbours of a node. In a social context, if person $Y$ is friends with both  $X$ and $Z$, one might expect some kind of link between $X$ and $Z$. The measurement of the connectedness of neighbours of nodes is the clustering coefficient. This is often averaged over all nodes and a single value is given as the clustering coefficient of the network~\cite{Dorogovtsev}. An explicit example is given in Figure~\ref{fig:clustering1}. The node labelled $i$ has $k_i$ neighbours (in this case $5$). Between these there could be a possible $k_i(k_i-1)/2$ undirected links of which $y$ exist. For this example, $3$ links exist out of the possible $10$ between the neighbours of node $i$. The clustering of node $i$ is given by
\begin{eqnarray}\label{eqn:clustering}
C_i&=& \frac{2y}{k_i(k_i-1)},
\end{eqnarray}
and would be $0.3$ for our example. The clustering nature of a network can also be expressed as the average over all nodes of degree $k$ giving a clustering distribution (or \emph{spectrum}) , $\bar{C}(k)$.

As noted by Klemm and Egu\'iluz, in a BA network, the average clustering of a given node is independent of its degree \cite{klemm:2002a}, in contrast to the findings of Fronczak \emph{et al.} \cite{Holyst}, and tends to zero in the large-size (thermodynamic) limit. A network model which does exhibit high clustering is that of Watts and Strogatz~\cite{smallworld} in which a random rewiring process is carried out on an initially regular lattice. The networks generated by such a process feature short average shortest path lengths between node pairs (the \emph{small-world} effect). However, these networks do not exhibit a power-law degree distribution. Subsequently, there have been many efforts to build models which can encompass both of these features such as the non-equlibrium growing network model proposed by Klemm and Egu\'iluz which features deactivation of a nodes availibility to be connected to \cite{klemm:2002a}. Other models such as that of  Holme and Kim \cite{Holme} and that introduced by  Toivonen \emph{et al.} \cite{toivonen:2006} modified the original BA preferential attachment mechanism, allowing further links between the new node and neighbours of the preferentially selected node. The social network model of Toivonen \emph{et al.} produced communities with dense internal connections.  Szab\'{o} \emph{et al.} formulated a scaling assumption and a mean-field theory of clustering in growing scale-free networks and applied it to the Holme and Kim mechanism \cite{Szab}. As discussed by Bogu\~n\'a and Pastor-Satorras, clustering in networks is closely related to degree correlations \cite{boguna:2003}. Infact, based on the work of Szab\'{o} \emph{et al.}, Barrat and Pastor-Satorras introduced a framework for computing the rate equation for two vertex correlations in the continuous degree and continuous time approximation \cite{Barrat1}. We shall now describe these degree correlations.

Vertex degree correlations are measures of the statistical dependence of the degrees of neighbouring vertices in a network. In general, $n$-vertex degree correlations, or $n$ point correlations, can be fully characterised by the conditional probability distribution 
\mbox{$P(k_1,k_2,$ $\ldots,k_n | k=n)$} that a vertex of degree $k=n$ is connected to a set of $n$ vertices with degrees $k_1,k_2, \ldots, k_n$. Two and three point correlations are of particular interest in complex networks as they can be related to network assortativity and clustering respectively.  More specifically, two vertex degree correlations (two point correlations) can be expressed as conditional probability $P(k'|k)$ that a vertex of degree $k$ is connected to a vertex of degree $k'$. Similarly, three vertex degree correlations (three point correlations) can be fully characterised by the conditional probability distribution $P(k', k''|k)$ that a vertex of degree $k$ is connected to both a vertex of degree $k'$ and a vertex of degree $k''$. This implies that the degrees of neighbouring nodes are not statistically independent. Reliable estimation of $P(k'|k)$ and $P(k', k''|k)$ requires a large amount of data and, in practice, one often resorts to related measures. Instead of $P(k'|k)$, the average degree of nearest neighbours, $\langle k_{nn}\rangle_k$, is often measured. This can be formally related to $P(k'|k)$ \cite{boguna:2003, pastor-satorras:2001}. If $\langle k_{nn}\rangle_k$ increases with $k$, high degree vertices tend to connect to high degree vertices. A network with this property would be described as being \emph{assortative} or displaying positive degree correlations. If $\langle k_{nn}\rangle_k$ decreases with $k$, high degree vertices tend to connect to low degree vertices (\emph{disassortative} or negatively correlated) \cite{boguna:2003, Newman}. An alternative to $\langle k_{nn}\rangle_k$ is to use a normalised Pearson's correlation coefficient of adjacent vertex degrees providing a single number measure of assortativity as suggested by Newman to further classify networks \cite{Newman, NewmanReview}. These approaches are discussed in more detail in Section~\ref{sec:degrees}. To characterise three point correlations, instead of using $P(k', k''|k)$, one can employ the clustering spectrum $\bar{C}(k)$, which can be related to $P(k',k''|k)$ \cite{boguna:2003}. In many real world networks such as the Internet~\cite{pastor-satorras:2001}, the clustering spectrum is a decreasing function of degree and while this is sometimes interpreted as a signature of hierarchical structure in a network, Soffer and V\'azquez suggested that this is a consequence of degree-degree (two point) correlations that enter the definition of the standard clustering coefficient \cite{soffer:2005}. The authors introduced a different definition for the clustering coefficient that does not have the degree-correlation `bias', i.e. a three-point correlation measure that filters out two-point correlations. Following the suggestion of Maslov {et al.} \cite{Maslov} that these phenomenon might arise from topological constraints rather than evolutionary mechanisms, Park and Newman demonstrated that dissasortative degree correlations observed in the Internet could be explained via the restriction of there being no double edges between nodes \cite{park:2003}. In contrast, social networks have been found to be assortative \cite{toivonen:2006}. Similarly, Catazaro \emph{et al.} observed that the network of scientific collaborations was assortative and presented a model to reproduce this feature \cite{caldarelli:2004}. 

%Other networks have been found to be assortative such as the collaboration patterns of scientists \cite{caldarelli:2004}. AND!!!!
%more here!!! Tiovenen, social network is positively assortative.

%other network models incorporating degree correlations
%Berg \emph{et al.} presented a model for the evolution of protein interaction networks based on two evolutionary processes and found that the imposed link dynamics, i.e. gain and loss of interactions through mutations in existing proteins, shaped the statistical structure of the network, resulting in correlations between the connectivities of interacting proteins in accordance with empirical findings \cite{berg:2004}. 

%correlations and function
%needs work!!
Functional processes occurring on networks are influenced by degree correlations, highlighting the importance of their role in complex networks. 
Egu\'iluz and Klemm considered highly clustered scale-free networks and showed that correlations play an important role in epidemic spreading \cite{EK}. The time average of the fraction of infected individuals in the steady-state undergoes a phase transition at a finite critical infection probability. They related this critical threshold to the transmission probability and the mean degree of nearest neighbours of all nodes in the system $\langle k_{nn}\rangle$, the conjectured criterion for epidemic spreading being related to the product of the two. The value $\langle k_{nn}\rangle$ scales with system size in their highly clustered scale-free network slower than in the random scale-free which is a byproduct of the dissasortativity of the system. Consequently, whereas in random scale-free networks in which viruses with extremely low spreading probabilities can prevail, the absence of connections between highly connected nodes in highly clustered scale-free networks protects the system against epidemics \cite{EK}. Interestingly, Bog\~un\'a \emph{et al.} assert that any scale-free network of appropriate exponent will have diverging $\langle k_{nn}\rangle$ in the thermodynamic (infinite size) limit, resulting in no threshold properties for epidemic spreading regardless of the correlations within the network \cite{boguna:absence} in contrast to the earlier suggestion of Bog\~un\'a and Pastor-Satorras~\cite{boguna:2002}. Brede and Sinha induced correlations to Erd\H{o}s-R\'enyi and scale-free networks by rewiring them appropriately to examine their dynamic stability \cite{brede:2005}. Each node had an associated variable whos evolution was governed by some non-linear function of the variables associated with neighbouring nodes. They mapped the adjacency matrices into Jacobian matrices and examined the largest eigenvectors, reflecting the decay rates of perturbations of the nodes' variables about their equilibrium states. They found that positive correlations within the network reduced the dynamical stability \cite{brede:2005}. Similarly, Bernardo \emph{et al.} induced negative degree correlations in scale-free networks whose links couple non-linear oscillators. Through analysis of the eigenratio (ratio of the eigenvalues) of the Laplacian of such networks, they found that network synchronisability improved as the network was made more dissassortative \cite{bernardo:2005a}. Based on this result, they conjectured that negative degree correlations may emerge spontaneously as the networked system attempts to become more stable \cite{bernardo:2005}. The same authors found similar results to hold also for weighted networks \cite{bernardo:2005a}. Maslov and Sneppen found that in regulatory networks, links between highly connected proteins were relatively rare increasing the robustness of the network to perturbations through ``localising deletous perturbations"  \cite{maslov:2002}. This is consistent with the findings of Berg \emph{et al.} whose model of the evolution of protein interaction networks exhibited disassortativity consistent with their empirical findings \cite{berg:2004}.

Krapivsky \emph{et al.} used a master equation method, in which rate equations for the densities of nodes of a given degree are employed, to investigate the steady state of the BA preferential attachment mechanism~\cite{KrapivskyDeriv}. This is a general method which can be applied to various growing network models. To extend the method to encompass two-point correlations, Krapivsky and Redner applied the approach to the number $N_{k,l}$ of nodes with total degree $k$ connected to ancestor nodes of total degree $l$ in a \emph{directed} network~\cite{Krapivsky}. They solved analytically the master equations for the steady-state of a specific directed network growth model, namely the Growing with Redirection algorithm~\cite{Krapivsky} which is further discussed in Section \ref{sec:Redner} and is similar to the mixture model introduced in Section~\ref{sec:model}. Bogu\~n\'a and Pastor-Satorras considered a link orientated description of two point correlations. For \emph{undirected} graphs, they introduced a symmetric matrix whose elements $E_{k,l}$ represent the number of links connecting nodes of degree $k$ to nodes of degree $l$~\cite{boguna:2003}. Bogu\~n\'a and Pastor-Satorras related this matrix to the joint degree distribution $P(k'\mid k)$ but noted that finite size effects make empirical evaluation of the matrix difficult, suggesting the use of the spectrum $\langle k_{nn} \rangle_k$ as a more suitable observable. Here, we introduce a similar matrix construction which can be applied to both the undirected \emph{and} directed scenarios. We call this the \emph{link-space} matrix. We show that it is possible to construct master equations to model the evolution of this matrix which can be applied to a wide variety of network evolution algorithms retaining important degree correlations which can be critical to the network's development. This framework is termed the \emph{link-space formalism} and is introduced in Section~\ref{sec:linkspace}. In certain cases, these master equations can be solved analytically providing a full time-dependent solution or a steady-state solution of the link-space matrix of the network. From these analytically derived link-space matrices both the degree and joint degree distributions can be obtained allowing accurate analysis of degree correlations. The formalism also allows the derivation of the form of networks of predetermined correlation properties such as a perfectly non-assortative network (Section~\ref{sec:degrees}) and the derivation of the steady-state solutions to network decay algorithms (Section~\ref{sec:decay}). The formalism can also be employed in its iterative guise to provide approximations to the steady-state when an analytic solution is not possible such as for the mixture model of Section~\ref{sec:model}.

One and two-point correlations have natural physical counterparts within the network, specifically the nodes and links. The one-point correlations $P(k)$ are simply related to the fraction of nodes in the network with degree $k$. As described in detail within this paper, two-point correlations $P(k'\mid k)$ are related to the fraction of links within a network connecting nodes of degree $k$ with nodes of degree $k'$, hence the term \emph{link-space}. One can extend the analysis to three-point correlations, $P(k',k''\mid k)$. This would be related to the number (fraction) of pairs of links within a network sharing a common node of degree $k$, the remaining link ends being connected to nodes of degrees $k'$ and $k''$ and all possible `open triangles' within the network would have to be considered. Clearly, the process can be extended to arbitrary $n$-point correlations although the physical interpretation of the appropriate measurable quantities will become increasingly obscure.

\section{Notation}\label{sec:notation}
Throughout the subsequent discussions, we shall use the following notation:
\begin{eqnarray}
N &=&\textrm{the number of nodes (vertices) within a network,}\nonumber\\
M &=& \textrm{the number of links (edges) within a network,}\nonumber\\
X_i &=&  \textrm{the number of nodes of degree $i$ within a network,}\nonumber\\
c_i &=&   \textrm{the fraction of nodes of degree $i$ within a network}\nonumber\\
{}&=& \frac{X_i}{N}, \nonumber\\
p_i &=& \textrm{the probability of selecting an individual node in a network which has degree $i$,}\nonumber\\
P_i &=&	\textrm{the probability of selecting any node in a network with degree $i$,}\nonumber\\
{} &=& X_i p_i  ~~\textrm{if selecting at random,}\nonumber\\
L_{i,j}&=& \textrm{the number of links from nodes of degree $i$ to nodes of degree $j$ for $i\ne j$,}\nonumber\\
L_{i,i}&=& \left\{ \begin{array}{l}
\textrm{twice the number of links between nodes of degree $i$ in the undirected graph,}\\
\textrm{the number of links between nodes of degree $i$ in the directed graph,}\end{array}\right. \nonumber\\
l_{i,j} &=&\frac{L_{i,j}}{M} ,\nonumber\\
\langle k_{nn}\rangle_i&=& \textrm{average degree of the neighbours of nodes of degree $i$,} \nonumber\\
\langle \frac{1}{k_{nn}}\rangle_i&=& \textrm{average of 1/degree of the neighbours of nodes of degree $i$,} \nonumber\\
\beta_i &=& \frac{1}{i~c_i}\langle \frac{1}{k_{nn}}\rangle_i ,\nonumber\\
\Theta_i&=& \textrm{the probability of attachment to a node of degree $i$ in the existing network,}\nonumber\\
m &=& \textrm{the number of links added per new node in a growing network.} \nonumber\\
{}&{}&{}
\end{eqnarray}
%\newpage
\section{Random walkers to generate networks}\label{sec:Saramaki}
The use of random walkers to generate networks was introduced by Saram\"aki and Kaski~\cite{Saramaki}. In their model, as in many non-equilibrium evolving networks, one new node is attached to the existing network at each timestep with $m$ undirected links. The simplest implementation of their 
growth algorithm attaches this new node to the existing network component with $m=2$ links as follows:\\
\begin{enumerate}
\item Select a node at random within the existing network.
\item Perform a two-step random walk.
\item Establish (undirected) links between the nodes arrived at within the existing network and the new node. 
\end{enumerate}
Note that a link is not established between the new node and the initial node of the random walk. The generalisation is clear for $m>2$ links per new node attached. To summarize, a random walk of length $m$ steps is performed
within the existing network from a randomly chosen start point, and a link between the new node and each of the nodes reached  
in the random walk is formed with the exception of the start point.\\
\begin{figure}
\begin{center}
\includegraphics[width=0.65\textwidth]{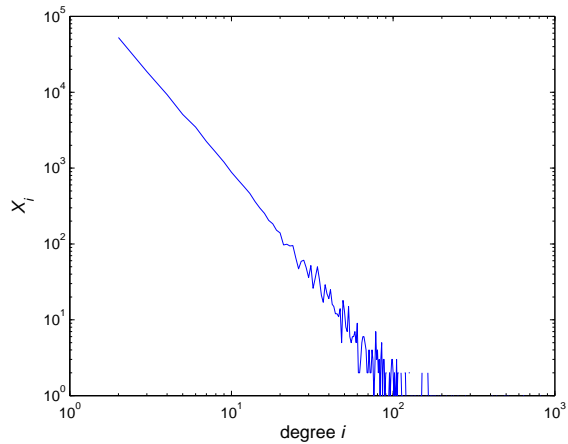}
\caption{ \label{fig:saramaki} The degree distribution of an example network using the Saram\"aki/Kaski algorithm grown to $10^5$
nodes.}
\end{center} 
\end{figure}

Certainly this yields results comparable to the scale-free preferential attachment of the BA model as shown in  
Figure~\ref{fig:saramaki}. However, the analytic approximations derived by Saram\"aki and Kaski in reference~\cite{Saramaki} do not hold in all cases. Their analysis is as follows. Consider choosing some vertex labeled $A$ initially (at random) within the existing network comprising $N$ nodes. This occurs with probability:
\begin{eqnarray}
p(A)&=&\frac{1}{N}.
\end{eqnarray}

We now wish to consider the probability of moving to a neighbour vertex labeled $B$ after one step of the random walk.  
Using Bayes theorem,
\begin{eqnarray}
p'(B)&=&\frac{p(B\mid A) p(A)}{p(A\mid B)},
\end{eqnarray}
where $p'(B)$ denotes arriving at node $B$ after one step, $p(B\mid A)$ denotes the probability of arriving at at $B$ if node  
$A$ is the initial vertex chosen and $p(A\mid B)$ is the conditional probability that, given we have arrived at $B$, we  
originated at $A$. 
The value $p(B\mid A)$ can be trivially written
\begin{eqnarray}
p(B\mid A) &=& \frac{1}{k_A},
\end{eqnarray}
where $k_A$ is the degree of vertex $A$. So far so good.
However, the next justification is somewhat misunderstood. It is often thought~\cite{Saramaki} that the conditional probability $p(A\mid B)$ can be  
written
\begin{eqnarray}
p(A\mid B) &=& \frac{1}{k_B},
\end{eqnarray}
which combined with the above yields:
\begin{eqnarray}\label{eqn:wrong}
p'(B)&=&\frac{k_B}{N k_A}.
\end{eqnarray}
This compares to the the BA preferential attachment probability for attaching to a particular node of degree $i$:
\begin{equation}
p_i = \frac{i}{\sum_j k_j},
\end{equation}
such that the probability of attaching to a node is proportional to its degree.
\begin{figure}
\begin{center}
\includegraphics[width=0.65\textwidth]{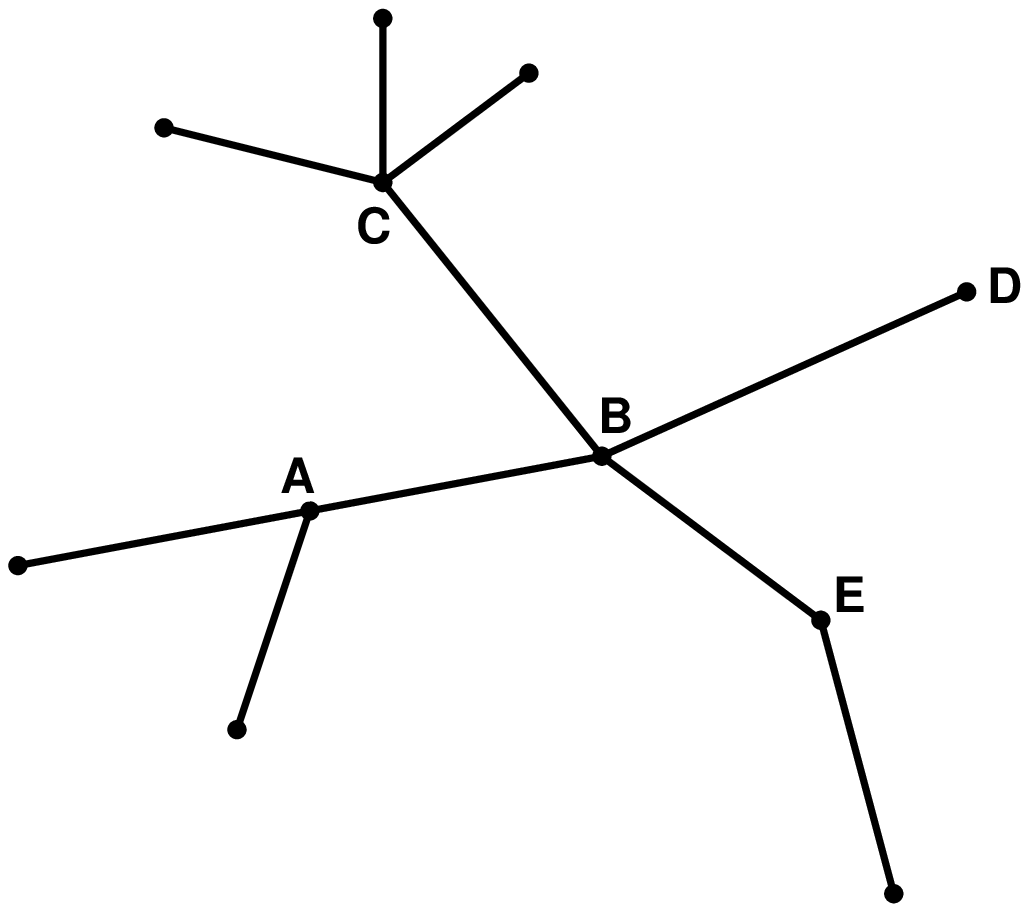}
\caption{ \label{fig:simplenet} A trivial example network.}
\end{center} 
\end{figure}

To understand the problems with this analysis, consider a real system of nodes, as described in Figure~\ref{fig:simplenet}. We can write the probability associated with arriving at node $B$ after a random walk of one step in terms of the initial node selected being one of the neighbours of $B$ and the subsequent probability of moving to node $B$:
\begin{eqnarray}
p'(B)&=& p(A)p(A\rightarrow B)+p(C)p(C\rightarrow B)+p(D)p(D\rightarrow B)+p(E)p(E\rightarrow B),\nonumber\\
{}&=& \frac{1}{N}\big\{\frac{1}{k_A}~+~\frac{1}{k_C}~+~\frac{1}{k_D}~+~\frac{1}{k_E}\big\},\nonumber\\
{}&=&\frac{25}{12 N}.
\end{eqnarray}
Clearly this is not the same as the earlier expression (\ref{eqn:wrong}). Nor should it be. The earlier expression would yield different results depending on which node was labeled $A$ which is clearly a nonsense.

Another interesting assumption often made is that because the initial node is  chosen  
at random, its vertex degree will be distributed similarly to that of the entire network~\cite{Saramaki}. If this is so, (\ref{eqn:wrong}) can be  written in terms of arriving at a particular of node of degree $i$ after randomly choosing an initial vertex and moving at random to one of its neighbours:
\begin{eqnarray}
p'_i &=& \frac{i}{N \langle k\rangle},
\end{eqnarray}
where $\langle k\rangle$ is the mean degree of the existing network component. Even averaging over all nodes of degree $i$, and (incorrectly) assuming that on average the neighbours of these nodes are distributed identically to the network as a whole, this would have implied the entirely different result
\begin{eqnarray}\label{eqn:picalc}
\langle p'_i\rangle &=& \frac{P'_i}{X_i} \nonumber\\
{}&=& \frac{i}{N} \Big< \frac{1}{k}\Big>.
\end{eqnarray}
We will discuss a similar expression in Section~\ref{sec:model} and the mean degree of nearest neighbours in Section~\ref{sec:degrees}.

\begin{figure}
\begin{center}
\includegraphics[width=0.65\textwidth]{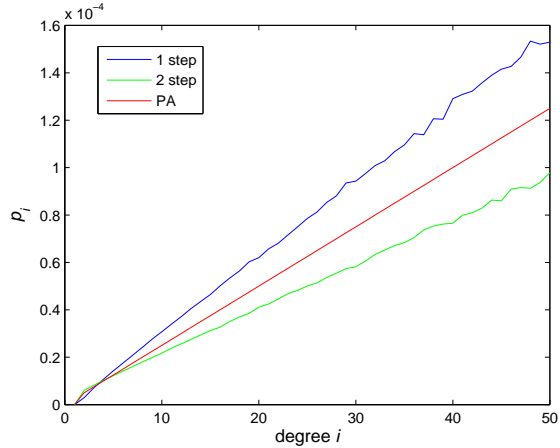}
\caption{ \label{fig:correlation1} The probability of arriving at a specific node of degree $i$ after a random walk of one and two steps compared to the preferential attachment probability. This analysis has been carried out on a network comprising $10^5$ nodes grown using the Saram\"aki and Kaski algorithm.}
\end{center} 
\end{figure}

In fact, the inherent degree-degree correlations are such that they cannot be ignored. We can see this by looking at an example network grown to $N = 10^5$ nodes using the simplest form ($m=2$) of the Saram\"aki algorithm outlined earlier. Having generated such a network, we shall investigate the probability of arriving at a node of degree $i$ after a random walk of one and two steps and compare these to the preferential attachment probability. Transforming the adjacency matrix of the network generated by the algorithm into a Markov-Chain transition matrix~\cite{MarkovChain} (dividing the columns by their sums) we can look at all possible random walks from all possible starting points for a number of steps. This is simply acheived by premultiplying a vector with all elements $1/N$ by the transition matrix. The result is a vector containing the probabilities of arriving at each node after a random walk of one step from random initial node. Multiplying this vector by the transition matrix again, provides the probabilities of arriving at each node in the network after a two-step random walk. Knowing the degrees of all the nodes in the network, it is then straight-forward to look at the probability of arriving at a degree $i$ node after a one or two step random walk. Whilst the degree distribution is described by $c_i$ which is the same as $P_i~=~\frac{X_i}{N}$ for selecting  nodes at random, the BA preferential attachment probability is described in terms of individual node probabilities $p_i$. We use the first two parts of (\ref{eqn:picalc}) to infer the probability of arriving at a specific node of degree $i$ after one or two steps. The preferential attachment probability would correspond to an infinite random walk. The comparison is illustrated in Figure~\ref{fig:correlation1}. Clearly, a one-step random walk is biased towards high degree nodes in comparison to preferential attachment whilst a two step random walk is biased towards low degree nodes.

\begin{figure}
\begin{center}
\includegraphics[width=0.65\textwidth]{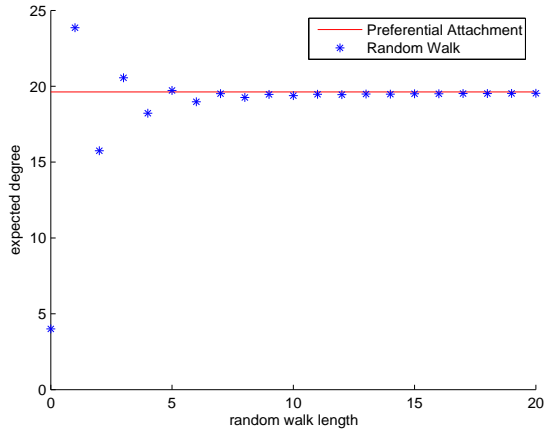}
\caption{ \label{fig:correlation2} The expected degree of the node arrived at after a number of steps in a random walk from  random initial node. This is performed on a realisation of the Saram\"aki-Kaski network, $m=2$ for $10^5$ nodes.}
\end{center} 
\end{figure}

To understand why this model yields the scale-free degree distributions equivalent to preferential attachment, we must look at the expected degree of the nodes in the existing network to which the new node is to be attached. Using the same transition matrix process, the mean degree of nodes reached after a number of steps of random walk can be calculated for a given network (here $10^5$ nodes) for all possible random walks from random starting positions. The expected degree of the first node in the walk is much higher than the convergent (preferential attachment) value, and the second lower as shown in Figure~\ref{fig:correlation2}. By interpolating between these two, an approach to preferential attachment has been achieved for this network growth algorithm, even for $m=2$ links added (2 step random walk) per new node.

We will make use of these correlations in our proposed model (see Section~\ref{sec:model}).

\section{Node-space and Link-space}\label{sec:linkspace}
We now introduce the link-space formalism which is based upon the master equation method. Consider a simple, growing, non-equilibrium network in which one node is added to a network at each timestep and this node is connected to the existing network with exactly $m$ undirected links. The process is governed by an attachment probability \emph{kernel} $\Theta_j$, defined as the probability that a specific, newly-introduced link attaches to any node of degree $j$ within the existing network. At some time $t$ there exist $X_i(t)$ nodes of degree $i$ and we wish to compute the expected number of nodes with degree $i$ at time $t+1$. The \emph{node-space master equations} can be expressed in terms of the attachment kernels and are written
\begin{eqnarray}
\langle X_i(t+1) \rangle  &=& X_i(t) +m\Theta_{i-1}(t)-m\Theta_{i}(t) \, \, \,i>m,\nonumber\\
\langle X_m(t+1)\rangle &=& X_m(t) + 1 - m\Theta_m(t),
\label{eqn:nodeXmaster}
\end{eqnarray}
since a new node of degree $m$ is added to the existing network at each timestep and there are no nodes with degree less than $m$. 

So far we have said nothing about the attachment mechanism, and have made
the easily geneneralisable restriction that only one node is being added per timestep with undirected links. We now follow a similar analysis, but retain the node-node linkage correlations that are inherent in many real-world systems \cite{Callaway, Krapivsky, NewmanReview}.
Consider any link in a general network -- we can describe it by the degrees of the two nodes that it connects. 
Hence we can construct a matrix $\mathbf{L}(t)$ such that the element $L_{i,j}(t)$ is equal to the number of links from nodes of degree $i$ to nodes of degree $j$ for $i\ne j$ at some time $t$. To ease the mathematical analysis below, the diagonal element $L_{i,i}(t)$ is defined to be \emph{twice} the number of links between nodes of degree $i$ for the undirected graph, a mathematical convienience also observed by Bogu\~n\'a and Pastor-Satorras~\cite{boguna:2003}. For undirected networks $\mathbf{L}(t)$ is symmetric and $\sum_{i,j} L_{i,j}(t) = 2M(t)$, twice the total number of links $M(t)$ in the network which is simply $mt$ when introducing $m$ links per timestep.  An explicit example is given for the simple network discussed in Section~\ref{sec:Saramaki} which is shown in Figure~\ref{fig:simplenet}. This network will also be further analysed in Section~\ref{sec:model}. The link-space matrix for this network is then expressed as in (\ref{eqn:explicit}). Note the double counting of the $4\leftrightarrow4$ link ($B \leftrightarrow C$).
 \begin{eqnarray}\label{eqn:explicit}
\mathbf{L}&=& \left( \begin{array}{cccc}
 0&1 & 2 &4\\
 1&0 & 0 &1\\
 2&0 & 0 &1\\
 4&1 & 1 &2
 \end{array} \right) .
 \end{eqnarray}
The matrix, $\mathbf{L}$, represents a surface describing degree-degree correlations and is called the \emph{link-space} matrix.

Consider one of the newly introduced links in an evolving network, one end of which is attached to the new node. The probability of selecting any node of degree $i-1$ within the existing network for the other end to attach to is given by the attachment probability $\Theta_{i-1}(t)$. Suppose an $i-1$ node is selected. The fraction of nodes of degree $i-1$ that are connected to nodes of degree $j$ is 
\begin{displaymath}
\frac{L_{i-1,j}(t)}{(i-1)X_{i-1}(t)}.
\end{displaymath}
The expected increase in links from nodes of degree $i$ to nodes of degree $j$, through the 
attachment of the new node to a node of degree $i-1$, is given by 
\begin{displaymath}
\frac{\Theta_{i-1}(t) L_{i-1,j}(t)} { X_{i-1}(t)}.
\end{displaymath}
Since each link has two ends, the value $L_{i,j}$ can increase by a connection to an $(i-1)$-degree node which is in turn connected to a $j$-degree node, or by connection to an $(j-1)$-degree node which is in turn connected to an $i$-degree node. We write master equations governing the evolution of the link-space matrix as the evolution of the expected number of links from $i$ to $j$ degree nodes, i.e. the number of $i \leftrightarrow j$ links. This is the \emph{link-space formalism} and is written for the (generalisable) case of adding one new node with $m$ undirected links to the existing network as
 \begin{eqnarray}
\langle L_{i,j}(t+1)\rangle  & =& L_{i,j}(t)  
+\frac{m\Theta_{i-1}(t)L_{i-1,j}(t)}{X_{i-1}(t)}\nonumber \\
{}&{}&+\frac{m\Theta_{j-1}(t)L_{i,j-1}(t)}{X_{j-1}(t)} -~\frac{m\Theta_{i}(t)L_{i,j}(t)}{X_{i}(t)}\nonumber \\
{}&{}& - \frac{m\Theta_{j}(t)L_{i,j}(t)}{X_{j}(t)}, \, \, \, i,j>m, \nonumber \\
\langle L_{m,j}(t+1)\rangle & =& L_{m,j}(t) + m\Theta_{j-1}(t)  +\frac{m\Theta_{j-1}(t)L_{m,j-1}(t)}{X_{j-1}(t)} \nonumber\\
{}&{} &- \frac{m\Theta_{m}(t)L_{m,j}(t)}{ X_{m}(t)}\nonumber \\
{}&{}&-\frac{m\Theta_{j}(t)L_{m,j}(t)}{ X_{j}(t)}, \, \, \,  j>m.
\label{eqn:lspace}
\end{eqnarray}

There are a variety of ways in which both the node-space and link-space master equations can be approached. For example, a full, time-dependent solution could be investigated as in Section~\ref{subsec:ER} or, using appropriate initial conditions, the equations can be iterated over the required timescale as in Section~\ref{sec:model}. We can also investigate the possibility of a steady state of the algorithm under scrutiny. To do so, we assume that there exists a steady state in which the degree distribution remains static\footnote{The master equation does not represent the evolution of an ensemble average of networks. Each specific realisation will have its own evolution of attachment kernel which cannot be described by the ensemble average. The existence of a steady state to the master equation does not necessarily imply that specific realisations converge upon it in the large $N$ limit but simply that the solution is static with respect to the growth algorithm. In practice, however, a master equation approach often yields good analytic agreement with even single realisations.} and investigate a solution. Under this assumption, the fraction of nodes $c_i(t) = X_i(t)/N(t)$ which have a given degree remains constant such that $\langle X_i(t+1) \rangle - X_i(t) \approx {dX_i}/{dt} = c_i $ when one new node is added per timestep ($N(t)=t$). It is also assumed that in the steady state, the attachment kernels are static too. We drop the notation `$(t)$' to indicate the steady-state and can rewrite (\ref{eqn:nodeXmaster}) as
\begin{eqnarray}
c_i &=&m\Theta_{i-1}-m\Theta_{i}, \,\,\, i>m,\nonumber\\
c_m &=&1-m\Theta_m.
\label{eqn:nsmaster}
\end{eqnarray}
The fraction of links between nodes of degree $i$ and nodes of degree $j$ can be expressed as the normalised link-space matrix, $l_{i,j}(t) = L_{i,j}(t)/M(t)$, which sums to $2$ in the undirected case. In the steady state we can assume that these values are static and can rewrite the link-space master equation (\ref{eqn:lspace}) as 
\begin{eqnarray}
l_{i,j}& =& \frac{\frac{\Theta_{i-1}}{c_{i-1}}l_{i-1,j} ~+~ 
\frac{\Theta_{j-1}}{c_{j-1}}l_{i,j-1}}{\frac{1}{m}~+ 
\frac{\Theta_{i}}{c_{i}}~+~\frac{\Theta_{j}}{c_{j}}  },\, \, \,i,j>m,  
\nonumber\\
l_{m,j}& =& \frac{\frac{\Theta_{j-1}}{c_{j-1}}l_{m,j-1}~+~\frac{\Theta_{j-1}}{m}}{\frac{1}{m}~+ 
\frac{\Theta_{m}}{c_m}~+~\frac{\Theta_{j}}{c_{j}} }, \, \, \,j>m.
\label{eqn:linkspacemaster}
\end{eqnarray}
The notation `$(t)$' has again been dropped to indicate the steady-state and here the generalisable situation of adding one new node per timestep is considered.

To apply the link-space formalism one starts by specifying the model dependent attachment kernel $\Theta_i$. To investigate a time-dependent solution, the attachment kernel is substituted into (\ref{eqn:nodeXmaster}) and (\ref{eqn:lspace}). To investigate a steady-state solution, the kernel is substituted into the (\ref{eqn:nsmaster}) and (\ref{eqn:linkspacemaster}) to yield recurrence relations for $c_i$ and $l_{i,j}$ respectively which can be solved analytically in some cases. The number of $i$-degree nodes is given by $X_i(t) = i^{-1} \sum_k L_{i,k}(t)$ which allows us to retrieve the degree distribution from the normalised link-space matrix:
\begin{eqnarray}\label{eqn:dist}
c_i(t) &=& \frac{M(t) \sum_{k=1}^{\infty} l_{i,k}(t)}{i N(t)}. 
\end{eqnarray}

Degree distributions of empirically observed or simulated networks typically become dominated by noise at large degrees reflecting the small probabilities associated with these values occurring. In the link-space, the situation is exacerbated and the high $i,j$ limit reflects connections between these high degree nodes. This is, of course, rarer than the existence of nodes of either degree. Following the conventional approach which is applied to degree distributions~\cite{Dorogovtsev}, we can use a cumulative representation of the link-space to address this issue. The use of a cumulative binning technique averages over stochasticity in the system. With degree distributions, regression techniques (curve fitting) applied to the cumulative distribution is used to obtain a more accurate description of the actual degree distribution than would be obtained from fitting to the empirical distribution itself~\cite{Dorogovtsev}. The process can be similarly applied in the link-space. Surface fitting could be applied to the cumulative link-space obtained from an empirical network. From the cumulative fitted surface, a more accurate representation of the actual link-space could be obtained for the empirical network, from which a better representation of the network's correlations could be obtained. This process also allows for comparison between simulated and analytically-derived link-space matrices. We define the cumulative link-space matrix, $_{cum}l_{i,j}$ to be
\begin{eqnarray}
_{cum}l_{i,j}&=&\sum_{x=i}^\infty \sum_{y=j}^\infty l_{x,y}.
\end{eqnarray}
Note that we have not lost generality in that given a cumulative linkspace matrix, the actual link-space can be derived using variations on the following:
\begin{eqnarray}
l_{i,j}&=&_{cum}l_{i,j} - _{cum}l_{i+1,j}-_{cum}l_{i,j+1}+_{cum}l_{i+1,j+1}.
\end{eqnarray}
The computation involved when evaluating $_{cum}l_{i,j}$ can be cut down considerably by first evaluating $_{cum}l_{1,1}$, which in the undirected graph is equal to $2$. The leftmost column (or top row) for $i>1$ can then be evaluated as
\begin{eqnarray}
_{cum}l_{i,1}&=& _{cum}l_{i-1,1}-\sum_{x=1}^\infty l_{i-1,x}.
\end{eqnarray}
Note that the second term on the right hand side is a row sum of the normalised link-space matrix and, hence, quickly evaluated. Indeed this row sum can be  related to the degree distribution $c_{i-1}$ from (\ref{eqn:dist}). Subsequent elements can be evaluated using the following simple identity,
\begin{eqnarray}
_{cum}l_{i,j}&=&_{cum}l_{i-1,j}+_{cum}l_{i,j-1}\nonumber \\
{}&{}& - _{cum}l_{i-1,j-1}+l_{i-1,j-1}.
\end{eqnarray}

We will now demonstrate the use of the link-space formalism to study a random attachment model and the Barab\'asi-Albert (BA) model using steady-state solutions and the classical Erd\H{o}s and R\'enyi random graph using a time-dependent solution.

% -------------------------------------------------------------------------------------------------------------------------------------
\subsection{Random attachment model (steady-state solution).}
\label{subsec:RA} 
\begin{figure}
\begin{center}
\includegraphics[width=0.7\textwidth]{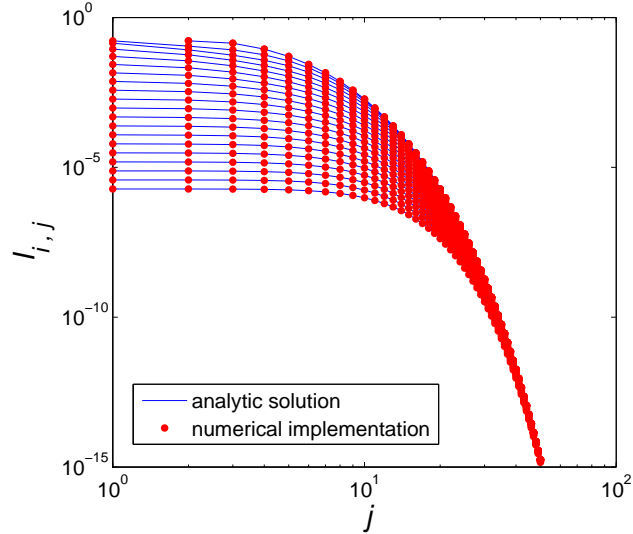}
\caption{ \label{fig:raanalvexact} Comparison of the numerically derived normalised link-space matrix and the analytic solution for random  attachment whereby one new node is added with one undirected link per timestep. The numerical implementation populates the link-space matrix directly from the degree distribution and the link-space recurrence relations. The first twenty rows ($i:1\to 20$) of the link-space matrices are illustrated.}
\end{center} 
\end{figure}

\begin{figure}
\begin{center}
\includegraphics[width=0.45\textwidth]{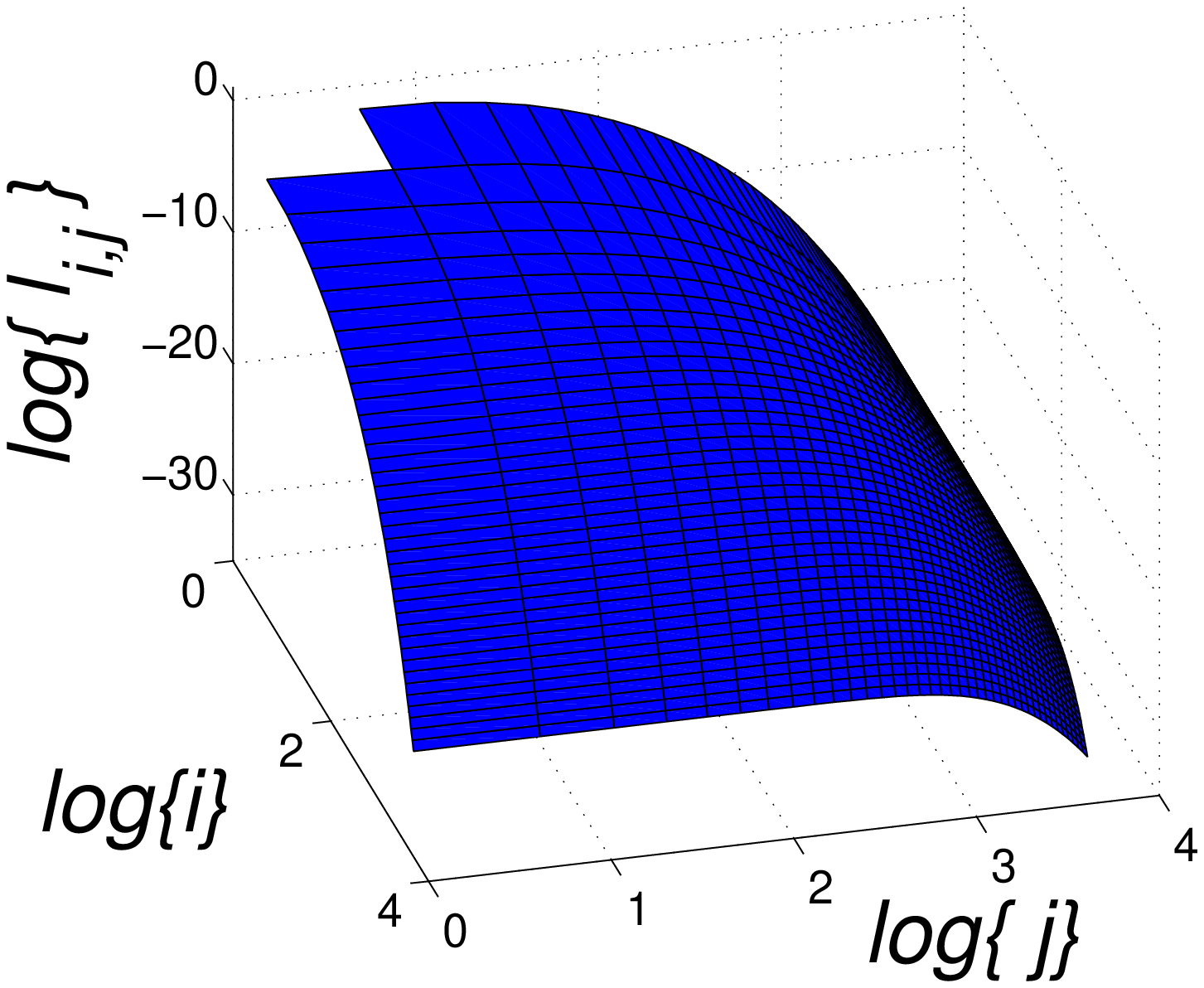}
\includegraphics[width=0.45\textwidth]{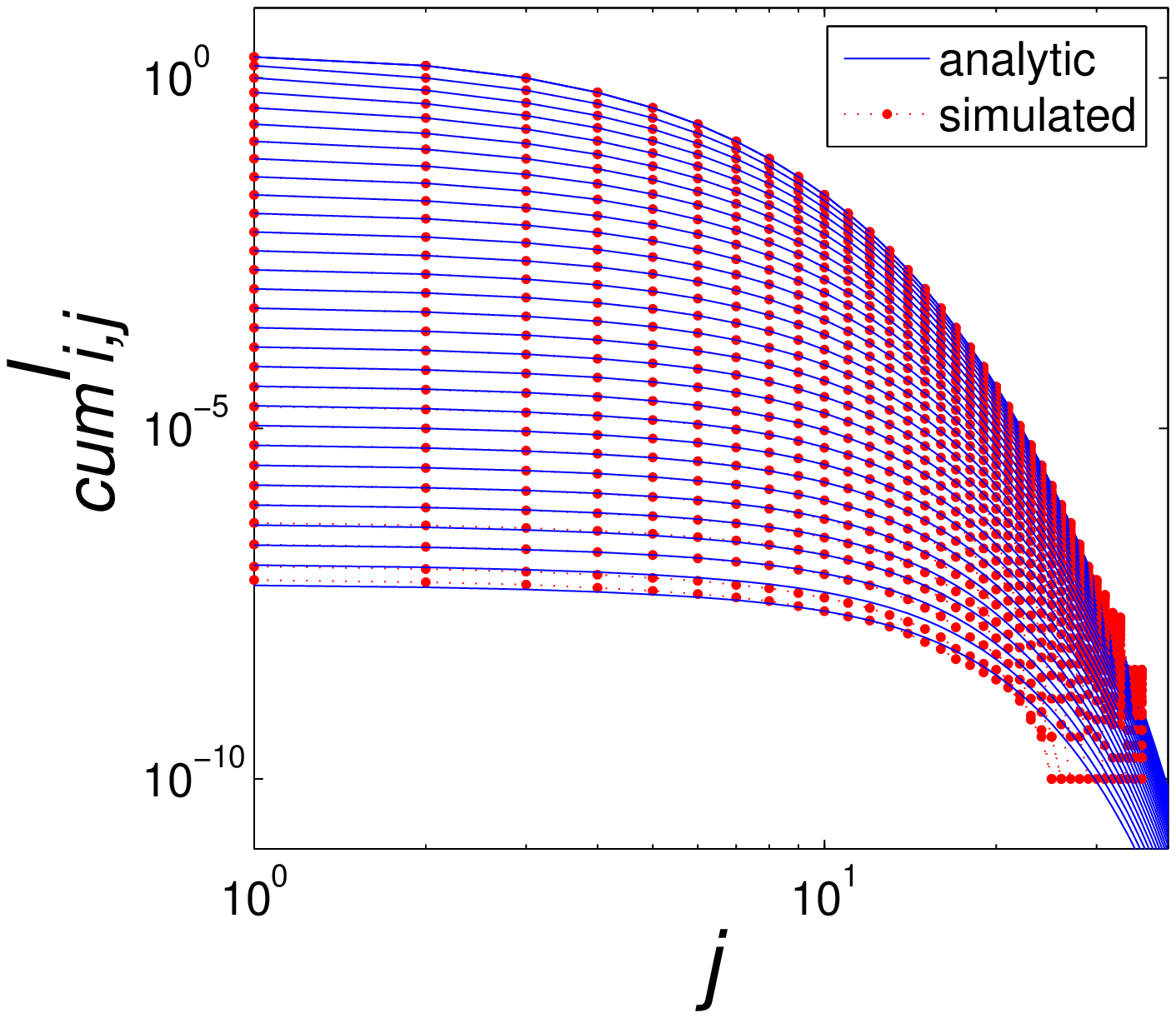}
\caption{ \label{fig:RA} Left: the analytically-derived, normalised link-space matrix for the random attachment growth algorithm. This could be filled to arbitrary size but is truncated to maximum degree of $40$ here. Right: comparison of the cumulative link-space matrices for the analytic solution and a simulation of the algorithm. The simulation comprises an ensemble average of $100$ networks grown to $10^8$ nodes. The maximum degree obtained was $37$. The first $30$ rows ($i$ values) of the cumulative link-space are illustrated and finite-size effects are noticable at high $j$. }
\end{center} 
\end{figure}

Consider first a random-attachment model in which at each timestep  a new node is added to the network and connected to an existing node with uniform probability without any preference (`Model A' in \cite{Barabasi,BarabasiDeriv2}) with one undirected link such that $m=1$ and $M(t)\approx N(t)= t$. We assume a steady-state solution and the attachment kernel is $\Theta_i= X_i/t =c_i$. Substituting into (\ref{eqn:nsmaster}), we obtain the recurrence relation $c_{i+1} = c_i / 2$, which yields the familiar degree distribution $c_i = 2^{-i}$. Substituting into (\ref{eqn:linkspacemaster}) yields the recurrence relation
 %DMDS: what's with the model a thing??
\begin{eqnarray}\label{eqn:ramaster2}
l_{i,j}&=& (l_{i-1,j}+l_{i,j-1}) / 3, \, \, \, \, i,j~>1,\nonumber\\
l_{1,j}&=& (c_{j-1}+l_{1,j-1}) / 3,\, \, \, \, j>1,
\end{eqnarray}
with $l_{1,1}=0$. It is easy to populate the link-space matrix numerically, just from the recurrence equations (\ref{eqn:ramaster2}) and the degree distribution. At first glance, the solution to the master equation (\ref{eqn:ramaster2}) would be of the form
\begin{eqnarray}
l_{i,j}&=& \frac{b}{4~2^{i+j}~3^{i+j}}.
\end{eqnarray}
However, the boundary conditions (which could be interpreted as influx of probability into the diffusive matrix) are such that this doesn't hold. 
 \begin{figure}
\begin{center}
 \includegraphics[width=0.6\textwidth]{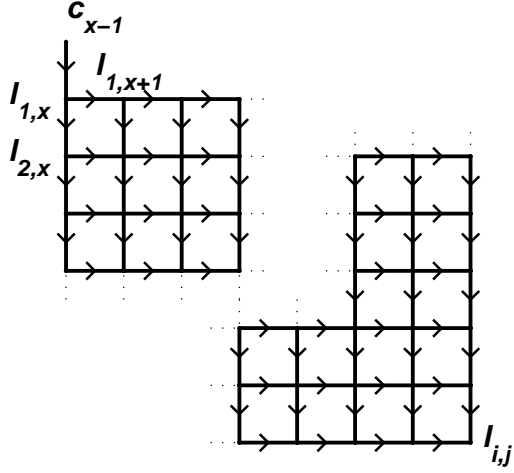}
\caption{ \label{fig:RAdiag} The paths of probability flux from $c_{x-1}$ influencing element $l_{i,j}$. Each arrow (step) represents a further factor of $\frac{1}{3}$.}
\end{center} 
\end{figure}
We can actually solve the link-space for this model exactly. Consider the values $c_i$ as being influxes of probability into the top and left of the link-space matrix. We can compute the effects of such influx on an element in the normalised link-space matrix, $l_{i,j}$. Each step in the path of probability flux reflects an extra factor of $\frac{1}{3}$. First, we will consider the influx effect from the top of the matrix as in Figure~\ref{fig:RAdiag}. 
The total path length from influx $c_{x-1}$ to element $l_{i,j}$ is simply $i+j-x$. The number of possible paths between $c_{x-1}$ and element  $l_{i,j}$ can be expressed ${j-x+i-1 \choose j-x}$ using the conventional binomial combinatorial or `choose' coefficient. We can similarly write down the paths and lengths for influxes into the left hand side of the matrix. The first row (and column) can be described as
\begin{eqnarray} 
l_{1,j}& =& \frac{c_{j-1}+l_{1,j-1}}{3}\nonumber\\
{}&=& \frac{2^{-(j-1)}+l_{1,j-1}}{3}\nonumber\\
{}&=& \sum_{k=1}^{j-1}\frac{1}{3^k 2^{j-k}}.
\end{eqnarray}
Subsequent rows can be similarly described. So, for  
$i,j>1$ an element can be written
\begin{eqnarray}
l_{i,j}&=& \frac{l_{i-1,j}+l_{i,j-1}}{3}\nonumber\\
{}&=&\sum_{x=2}^{j} \frac{  {i-1+j-x \choose j-x}} {3^{(i+j-x)}2^{(x-1)}}~+~ \nonumber\\
{}&{}&\sum_{x=2}^{i} \frac{{i-1+j-x\choose i-x }}{3^{(i+j-x)}2^{(x-1)}}. 
\end{eqnarray}
This expression is compared to a numerically populated link-space matrix in Figure~\ref{fig:raanalvexact}. 
We shall make use of this solution to investigate the correlations of such a network in Section \ref{sec:degrees}. This normalised link-space matrix is illustrated in Fig.~\ref{fig:RA} along with a comparison to simulated networks using the cumulative link-space matrix. 

% -------------------------------------------------------------------------------------------------------------------------------------
\subsection{Barab\'asi-Albert (BA) model (steady-state solution).}
\label{subsec:BA}
In the BA model \cite{Barabasi,BarabasiDeriv2}, at each timestep a new node is added to the network and connected to $m$ existing nodes with probabilities  proportional to the degrees of those nodes, i.e. $\Theta_i \propto i$ yielding 
\begin{equation}
\label{eqn:SF}
\Theta_i(t) = \frac{i X_i(t)}{\sum_{j} j X_j(t)} = \frac{i X_i(t)}{2M(t)} \approx \frac{i X_i(t)}{2mt}. 
\end{equation}
We consider the scenario when the new node is added with one undirected link, $m=1$, and the attachment kernel is well approximated by 
$\Theta_i \approx  i c_i / 2$. 
Substituting this into (\ref{eqn:nsmaster}) yields the recurrence relation 
$c_{i}=\frac{(i-1) c_{i-1}}{2}-\frac{i c_i}{2}
=\frac{i-1}{i+2}c_{i-1}$,
whose solution is $c_{i}=\frac{4}{i(i+1)(i+2)}$. This corresponds well to an actual network grown using this algorithm. Using the same substitution, the link-space master equations yield the recurrence relations:
\begin{eqnarray}\label{eqn:sfmaster3}
l_{i,j}&=& \frac{(i-1)l_{i-1,j}+(j-1)l_{i,j-1}}{2+i+j},\ \ \ i,j>1,\nonumber\\
l_{1,j}&=&\frac{(j-1)c_{j-1}+(j-1)l_{1,j-1}}{3+j},\ \ \ j>1.
\end{eqnarray}
Again, it is easy to populate the matrix numerically just by implementing the known degree distribution and the link-space recurrence equations but we can solve the link-space analytically. At first glance, the solution to the master equation (\ref{eqn:sfmaster3}) would be of the form
\begin{eqnarray}
l_{i,j}&=& \frac{w}{i(i+1)j(j+1)},
\end{eqnarray}
where $w$ is a constant. This would imply for the degree distribution
\begin{eqnarray}
c_i &=& \frac{w}{i^2(i+1)}.
\end{eqnarray}
Summing over the entire node-space would give $w = \frac{6}{\pi^2-6}$. However, the boundary conditions (which could be interpreted as influx of probability into the diffusive matrix) are such that this solution doesn't hold. We can obtain an exact (although somewhat less pretty) solution by tracing fluxes of probability around the matrix and making use of the previously derived degree distribution. The steady-state link-space master equation for general attachment kernel is written
\begin{eqnarray}
l_{i,j}& =& \frac{\frac{\Theta_{i-1}}{c_{i-1}}l_{i-1,j} +\frac{\Theta_{j-1}}{c_{j-1}}l_{i,j-1}}{1~+ \frac{\Theta_{i}}{c_{i}}~+~\frac{\Theta_{j}}{c_{j}}  } ~~~i,j>1  ,\nonumber\\
l_{1,j}& =& \frac{\frac{\Theta_{j-1}}{c_{j-1}}l_{1,j-1}+\Theta_{j-1}}{1~+ \frac{\Theta_{1}}{c_{1}}~+~\frac{\Theta_{j}}{c_{j}} } ~~~~~~~~j>1 ,\nonumber\\
l_{1,1}&=&0  .
\label{eqn:linkspacemastersf}
\end{eqnarray}
We can rewrite (\ref{eqn:linkspacemastersf}) in terms of vertical and horizontal components:
\begin{eqnarray}
l_{i,j}&=&\Psi_{i,j}l_{i-1,j} + \Upsilon_{i,j}l_{i,j-1} ,\nonumber\\
l_{1,j}&=& \Upsilon_{i,j}\big(l_{1,j-1} + c_{j-1}\big),\nonumber\\
l_{1,1}&=&0 ,
\end{eqnarray}
where
\begin{eqnarray}
\Psi_{i,j} &=& \frac{\frac{\Theta_{i-1}}{c_{i-1}}}{1+\frac{\Theta_{i}}{c_{i}}+\frac{\Theta_{j}}{c_{j}}},\nonumber\\
\Upsilon_{i,j} &=& \frac{\frac{\Theta_{j-1}}{c_{j-1}}}{1+\frac{\Theta_{i}}{c_{i}}+\frac{\Theta_{j}}{c_{j}}}.
\end{eqnarray}

\begin{figure}
\begin{center}
 \includegraphics[width=0.8\textwidth]{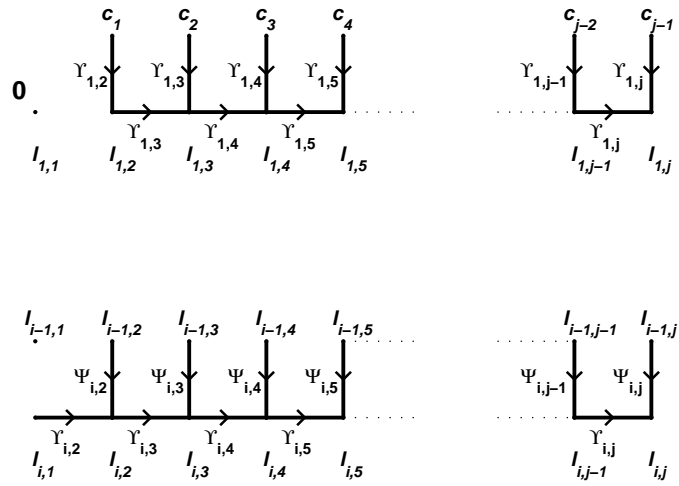}
\caption{ \label{fig:SFexact} The components of flux of probability around the normalised link-space matrix. This is general to any attachment kernel. An element within the link-space matrix can be built up from contributing elements and the appropriate factors.}
\end{center} 
\end{figure}
By considering the probability fluxes as shown in Figure~\ref{fig:SFexact}, we can write the individual elements in the link-space matrix as
\begin{eqnarray}
l_{1,j}&=&\sum_{y=2}^j \bigg(c_{y-1} \prod_{x=y}^j \Upsilon_{1,x}\bigg), \nonumber\\
l_{i,j}&=&\sum_{y=2}^j \bigg(l_{i-1,y} \Psi_{i,y} \prod_{x=y+1}^j \Upsilon_{i,x}\bigg) \nonumber\\
{}&{}&~+~l_{i,1}\prod_{x=2}^j \Upsilon_{i,x}, \nonumber\\
l_{1,1} &= &0.
\label{eqn:elements}
\end{eqnarray}
Note that we have yet to introduce the attachment probability kernels and the analysis so far is general.
Using the preferential attachment probability,
\begin{eqnarray}
\Theta_i & \approx & \frac{i c_i}{2} ,
\label{eqn:SFkernel3}
\end{eqnarray}
we can write our component-wise factors for the master equation
\begin{eqnarray}
\Psi_{i,j} &=& \frac{i-1}{i+j+2},\nonumber\\
\Upsilon_{i,j} &=& \frac{j-1}{i+j+2}.
\label{eqn:components}
\end{eqnarray}
Substituting (\ref{eqn:components}) into (\ref{eqn:elements}) and using the previously derived degree distribution yields for the first row
\begin{eqnarray}
l_{1,j}&=&\sum_{y=2}^j \bigg( \frac{4}{y(y-1)(y+1)} \prod_{x=y}^{j}\frac{x-1}{x+3}\bigg) \nonumber\\
{}&=&\frac{4(j-1)!}{(j+3)!}\sum_{y=2}^{j}(y+2) \nonumber\\
{}&=& \frac{2(j+6)(j-1)}{j(j+1)(j+2)(j+3)}.
\end{eqnarray}
Subsequent rows can be described:
\begin{eqnarray}
l_{i,j} &=&l_{i,1}\frac{(j-1)!(3+i)!}{(2+i+j)!}~+~\sum_{x=2}^j l_{i-1,x}(i-1)\frac{(j-1)!}{(2+i+j)!}\frac{(1+i+x)!}{(x-1)!}\nonumber\\
{}&=& \frac{(j-1)!}{(2+i+j)!}\bigg\{(3+i)!l_{i,1}~+~(i-1)\sum_{x=2}^j l_{i-1,x} \frac{(1+i+x)!}{(x-1)!}
\bigg\}.\label{eqn:working1}
\end{eqnarray}
Rewriting (\ref{eqn:working1}) gives
\begin{eqnarray}
l_{i,j}&=& \frac{(j-1)!}{(2+i+j)!}\bigg\{K_i~+~E_{i-1,j}\bigg\},
\label{eqn:working2}
\end{eqnarray}
where the meaning of $K_{i}$ and $E_{i-1,j}$ is transparent. Clearly, we can write (\ref{eqn:working2}) for $l_{i-1,x}$,
\begin{eqnarray}
l_{i-1,x}&=&\frac{(x-1)!}{(1+i+x)!}\bigg\{K_{i-1}~+~E_{i-1,x}\bigg\}.
\label{eqn:working3}
\end{eqnarray}
Substituting (\ref{eqn:working3}) into (\ref{eqn:working1}) yields
\begin{eqnarray}
l_{i,j}&=&  \frac{(j-1)!}{(2+i+j)!}\bigg\{K_i~+~(i-1)\sum_{x=2}^j\big(K_{i-1}~+~ E_{i-1,x}\big)\bigg\}.
\label{eqn:working4}
\end{eqnarray}
In order to solve this recurrence relation we define an operator for repeated summation, $S^n_{j,x}$ such that 
\begin{eqnarray}
S_{j,x}(f(x)) &=& \sum_{x=2}^j f(x),   \nonumber\\
S^n_{j,x}(f(x)) &=& \sum_{x_n=2}^j \sum_{x_{n-1}=2}^{x_{n}} \sum_{x_{n-2}=2}^{x_{n-1}}\dots
\sum_{x_3=2}^{x_4} \sum_{x_{2}=2}^{x_{3}}\sum_{x=2}^{x_2}f(x).
\end{eqnarray}
The subscripts denote the initial variable to be summed over and the final limit. 
A few examples of this operation clarify its use:
\begin{eqnarray}
S^0_{j,x}(f(x))& =&f(x),\nonumber\\ 
S_{j,x}(1)&=& j-1  ,\nonumber\\
S^2_{j,x}(1) &=& S_{j,x}(x)-S_{j,x}(1)\nonumber\\
{} &=& \frac{1}{2}(j^2-j),\nonumber\\
S^3_{j,x}(1)&=&\frac{1}{6}(j^3-j),\nonumber\\
\textrm{etc.}&{}&{}\nonumber\\ 
S_{j,x}(x)&=& \frac{j(j+1)}{2}-1 ,\nonumber\\
S^2_{j,x}(x) &=& \frac{1}{6}(j^3+3j^2-4j),\nonumber\\
\textrm{etc.}&{}&{}
\end{eqnarray}
We can use this operator in our expression for the element $l_{i,j}$ in (\ref{eqn:working4}) and expand to the value $E_{2,x}$,
\begin{eqnarray}
l_{i,j}&=& \frac{(j-1)!}{(2+i+j)!}\bigg\{ K_i~+~(i-1)S_{j,x}\big(K_{i-1}+E_{i-1,x}\big)\bigg\}\nonumber\\
{}&=& \frac{(j-1)!}{(2+i+j)!}\bigg\{ K_i+(i-1)K_{i-1}S_{j,x}(1) + (i-1)(i-2)S^2_{j,x}\big(K_{i-2} +E_{i-2,x} \big) \bigg\}   \nonumber\\
{}&=& \frac{(j-1)!}{(2+i+j)!}\bigg\{ K_i+(i-1)K_{i-1}S_{j,x}(1) + (i-1)(i-2)K_{i-2}S_{j,x}^2(1)\nonumber\\
{}&{}&+(i-1)(i-2)(i-3)K_{i-3}S_{j,x}^3(1)+...+(i-1)(i-2)(i-3)*...*2 K_2 S_{j,x}^{i-2}(1)\nonumber\\
{}&{}&  +(i-1)(i-2)*....*2S_{j,x}^{i-2}(E_{2,x}) \bigg\}.
\end{eqnarray}
We can express $E_{2,x}$ in terms of the operator $S$ too,
\begin{eqnarray}
E_{2,x} = 4 S_{x,x}\big(2S_{x,x}(1)+S_{x,x}(x)  \big).
\end{eqnarray}
The element $l_{i,j}$ can be expressed as
\begin{eqnarray}
l_{i,j}&=& \frac{(j-1)!}{(2+i+j)!}\bigg\{ \sum_{y=2}^i \frac{(i-1)!}{(y-1)!}K_y S^{i-y}_{j,x}(1) + 4 (i-1)!\big((2 S^i_{j,x}(1)+S^i_{j,x}(x)\big)\bigg\}.\nonumber\\
{}&{}&{}
\label{eqn:working6}
\end{eqnarray}
This form is somewhat obscure as the calculation of the operator values is less than obvious.
However, we can transform to a more easily interpreted operator $W(n)$ analogous to $S$ but with different limits (which enables further analysis) such that  
\begin{eqnarray}
W_{j,x}(f(x)) &=& \sum_{x=1}^j f(x),\nonumber\\
W^n_{j,x}(f(x)) &=& \sum_{x_n=1}^j \sum_{x_{n-1}=1}^{x_{n}} \sum_{x_{n-2}=1}^{x_{n-1}}\dots
\sum_{x_3=1}^{x_4} \sum_{x_{2}=1}^{x_{3}}\sum_{x=1}^{x_2}f(x).
\end{eqnarray}
Whilst, at first glance, it looks like little progress has been made, for this analysis we only need evaluate the repeated operation on initial function $f(x)=1$.  This is exactly solvable and as such, we can rewrite in terms of a function $G(n)$ dropping the superfluous $x$ subscripts,
\begin{eqnarray}
\label{eqn:inductive}
G_j(n)&=& W^n_{j}(1) ,\nonumber\\
G_j(n)&=& \left\{\begin{array}{l r}
\frac{(j+n-1)!}{n!(j-1)!}&\textrm{for $n\ge0$,}\\
0 & \textrm{for  $n<0$.} \end{array}\right.
\end{eqnarray}
The inductive proof associated with (\ref{eqn:inductive}) can be understood by path counting for some repeating binomial process and is an intrinsic property of the combinatorial choose coefficient. Consider a repeated coin toss over $x$ steps. The number of ways of achieving $n+1$ successes after these $x$ iterations is ${x \choose n+1}$. Now, the occurrence of this last success could have happened on the $(n+1)$th iteration or the following one, or any of the subsequent iterations till the $x$th one. For the last successful outcome to occur on the $m$th step, $n$ successful outcomes must have occurred in the previous steps. The number of ways this could have occurred is ${m-1 \choose n}$. Clearly, summing over all possible $m$ values, the total possible paths resulting in $n+1$ successes must equate to ${x \choose n+1}$. 

\begin{figure}
\begin{center}
\includegraphics[width=0.7\textwidth]{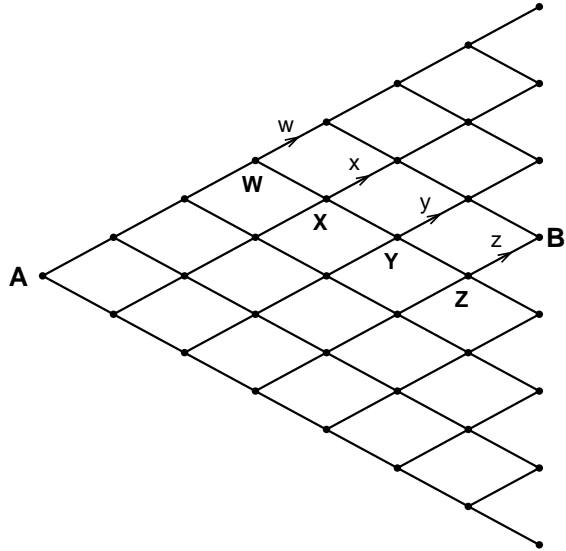}
\caption{ \label{fig:choosediag} A binomial process over seven steps. The number of paths between $\bf{A}$ and $\bf{B}$ can be expressed as the sum of the paths $\bf{A}\rightarrow\bf{W}$, $\bf{A}\rightarrow\bf{X}$, $\bf{A}\rightarrow\bf{Y}$ and $\bf{A}\rightarrow\bf{Z}$ .}
\end{center} 
\end{figure}

For clarity, this is depicted in Figure~\ref{fig:choosediag}. To  reach point $\bf{B}$ from $\bf{A}$ in the binomial process, one of the steps $w,x,y$ or $z$ must be traversed, after which there is only one route to $\bf{B}$. The number of paths between $\bf{A}$ and $\bf{B}$ utilising step $w$ is the same as the number of paths between $\bf{A}$ and $\bf{W}$. Similarly, the number of paths between $\bf{A}$ and $\bf{B}$ utilising step $x$ is the same as the number of paths between $\bf{A}$ and $\bf{X}$ and so on. The number of paths between $\bf{A}$ and $\bf{B}$ can be expressed as the sum of the paths $\bf{A}\rightarrow\bf{W}$, $\bf{A}\rightarrow\bf{X}$. $\bf{A}\rightarrow\bf{Y}$ and $\bf{A}\rightarrow\bf{Z}$. 
We incorporate this behaviour into our proof for the solution of $G_j(n)$,
\begin{eqnarray} 
G_j(n) &=& {j+n-1 \choose n} \nonumber\\
G_j(n+1)&=&\sum_1^j G_j(n)\nonumber\\
{}&=&\sum_1^j~{j+n-1 \choose n}\nonumber\\
{}&=&{j+n \choose n+1}
\end{eqnarray}
As $G(1)={j \choose 1}=j$, this inductive proof holds for all $n$ and the following relations hold
\begin{eqnarray}
S^n(1)&=& G(n)-G(n-1),\nonumber\\
S^n(j)&=& G(n+1)-G(n-1).
\end{eqnarray}

\begin{figure}
\begin{center}
\includegraphics[width=0.7\textwidth]{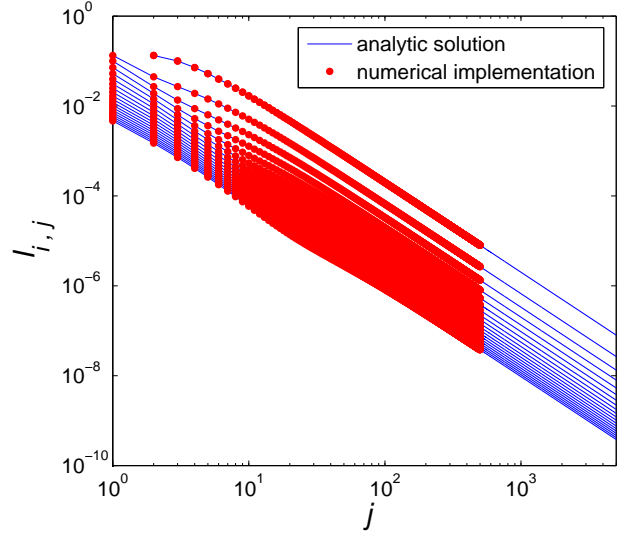}
\caption{ \label{fig:sfanalvexact} Comparison of the numerically derived normalised link-space matrix and the analytic solution for the steady state for  preferential attachment whereby one new node is added with one new link. The first twenty rows ($i:1\to 20$) of the link-space matrices are illustrated.}
\end{center} 
\end{figure}

\begin{figure}
\begin{center}
\includegraphics[width=0.45\textwidth]{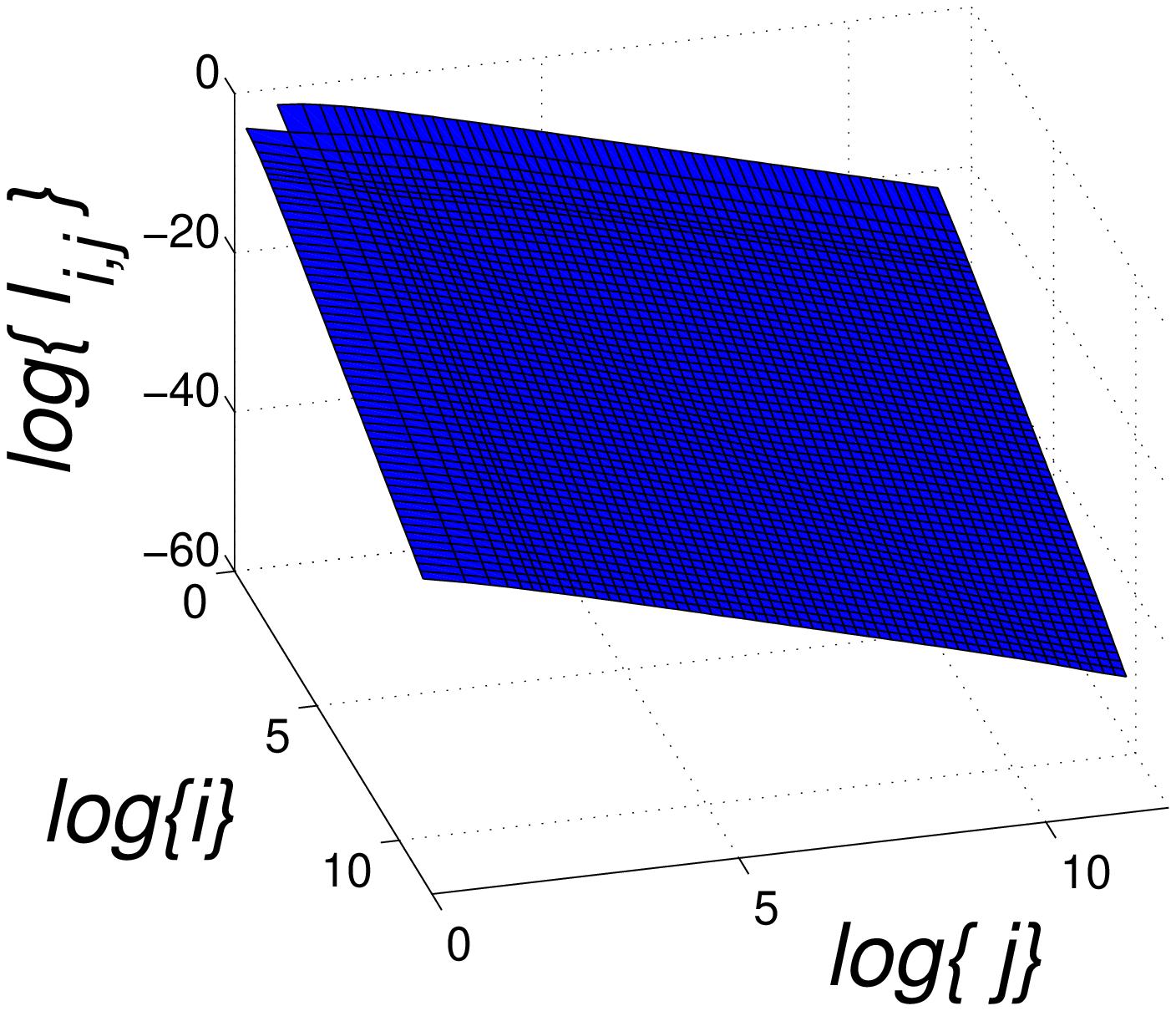}
\includegraphics[width=0.45\textwidth]{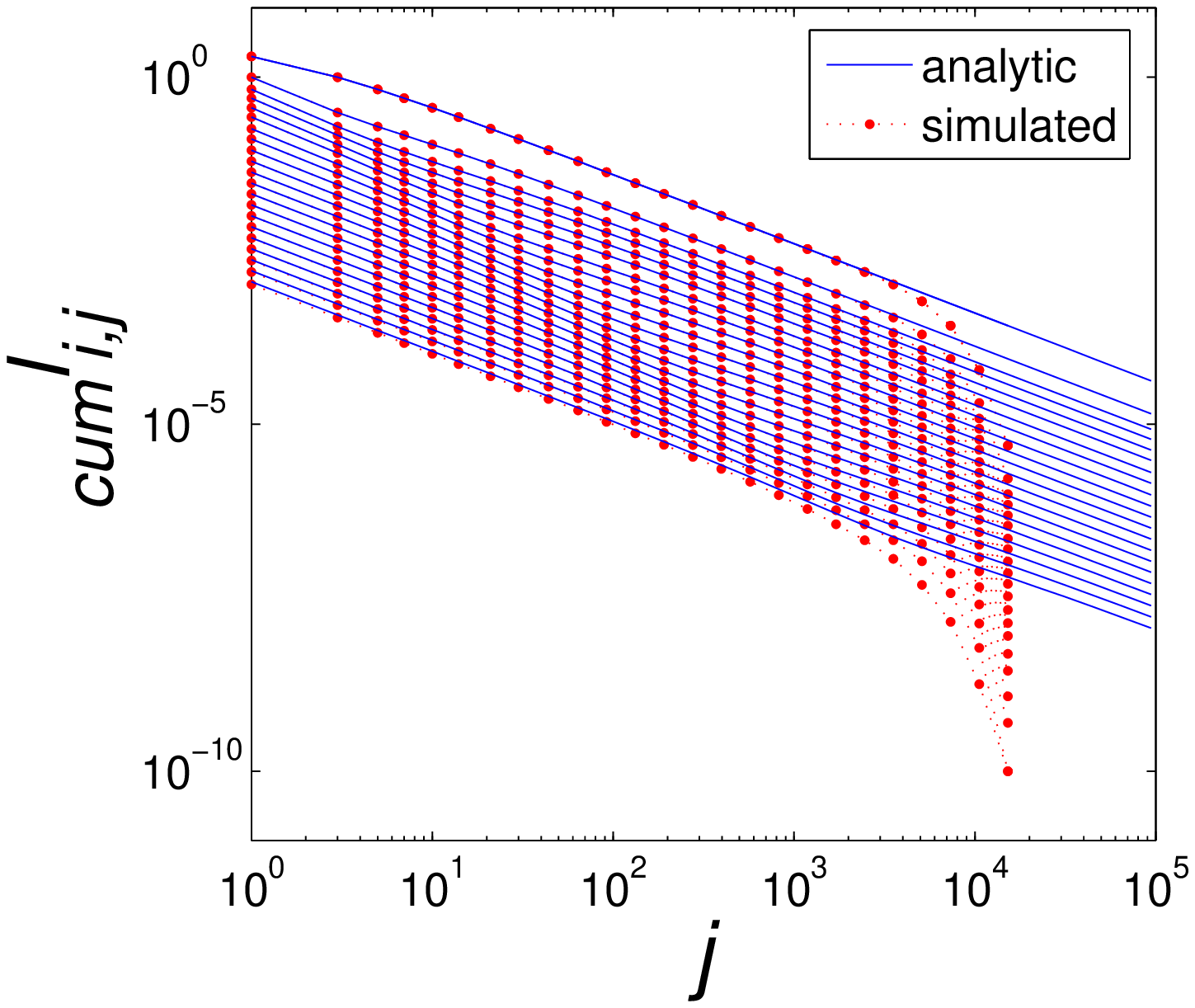}
\caption{ \label{fig:BA} Left: the analytically-derived, normalised link-space matrix for the preferential attachment growth algorithm. This could be filled to arbitrary size but is here truncated to maximum degree of approximately $10^5$. Right: comparison of the cumulative link-space matrices for the analytic solution and a simulation of the algorithm. The simulation comprises an ensemble average of $1000$ networks each grown to $10^7$ nodes. The maximum degree obtained was $17609$. Various rows ($i$ values) from $1$ to $3500$ of the cumulative link-space are illustrated and the effects of the finite nature of the simulation are apparent. }
\end{center} 
\end{figure}

We can now write the element in our link-space matrix for preferential attachment exactly as our function $G(n)$ is easily evaluated,
\begin{eqnarray}
l_{i,j}& =& \frac{4(j-1)!(i-1)!}{(j+i+2)!}\Bigg(G(i+1)+2G(i)-3G(i-1) \nonumber\\ {}&{}&+~\frac{1}{2}\sum_{k=1}^{i}(k-1)(k+6)\Big(G(i-k)-G(i-k-1)\Big)\Bigg), \nonumber\\
\end{eqnarray}
and the first few rows of our link-space can be written
\begin{eqnarray}
l_{1,j}& =& \frac{2(6+j)(j-1)}{j(j+1)(j+2)(j+3)},\\
l_{2,j}&=&\frac{2j(j-1)(j+10)+48}{3j(j+1)(j+2)(j+3)(j+4)}.
\label{eqn:sfexact}
\end{eqnarray}
A comparison of the numerically populated link-space matrix and the analytic solution for the BA preferential attachment model with $m=1$ is illustrated in Figure~\ref{fig:sfanalvexact}.
We shall use this solution to investigate the correlations with this network in Section~\ref{sec:degrees}. This normalised link-space matrix is illustrated in Fig.~\ref{fig:BA} along with a comparison to simulated networks. 

\subsection{Erd\H{o}s and R\'enyi random graph (time-dependent solution).}
\label{subsec:ER}
The Erd\H{o}s and R\'enyi (ER) classical random graph is the quintessential equilibrium network model~\cite{Dorogovtsev, ER}. However, to employ the link-space formalism, which tracks the evolution of a network's correlation properties, we must model it as an evolving, non-equilibrium network. The classical random graph can be modelled as a constantly accelerating network with random attachment. At each timestep, we add one new node to the existing network. All possible links between the new node and \emph{all} existing nodes are considered and each is established with probability $\alpha$. That is, a biased coin toss (Bernoulli trial) is employed for every node within the existing network to decide whether a link is formed between it and the new node. Conseqently the expected number of new, undirected links with which the new node connects to the existing network is $\langle m(t) \rangle  = \alpha N(t-1) \approx \alpha t$ and this would be described as an accelerating network~\cite{AcceleratingNetworks}. No new links are formed between existing nodes. This model subsequently produces a network in which the probability of a link existing between any pair of nodes is simply $\alpha$ and is thus representative of the ER random graph~\cite{ER}. A network grown to $t$ nodes will have mean degree $ \alpha t$ which will also be the expected degree of all nodes in the network. Whilst the random attachment and preferential attachment models both have steady-state assymptotic behaviour, clearly this model doesn't and a full, time-dependent solution is required.

\begin{figure}
\begin{center}
\includegraphics[width=0.45\textwidth]{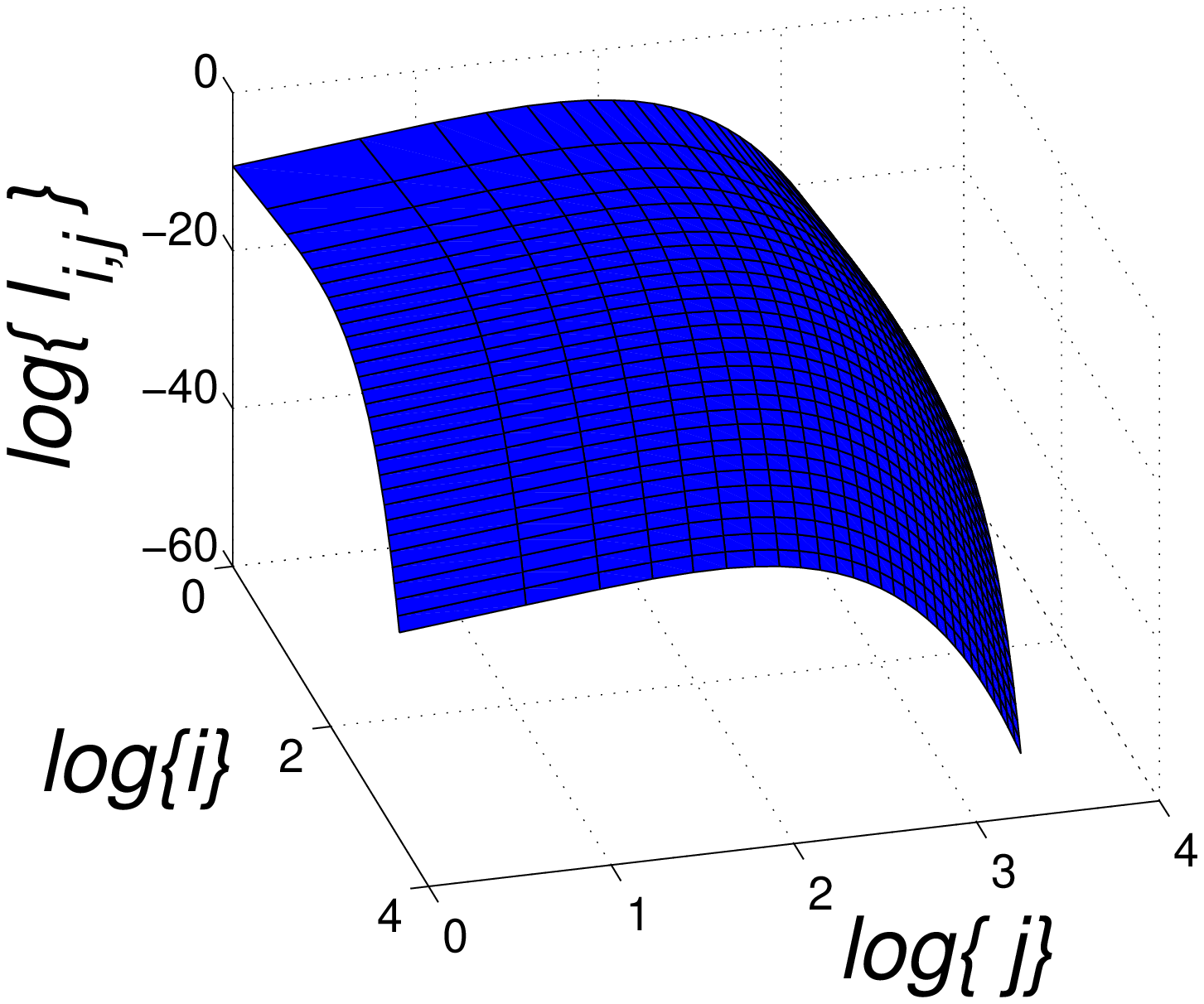}
\includegraphics[width=0.45\textwidth]{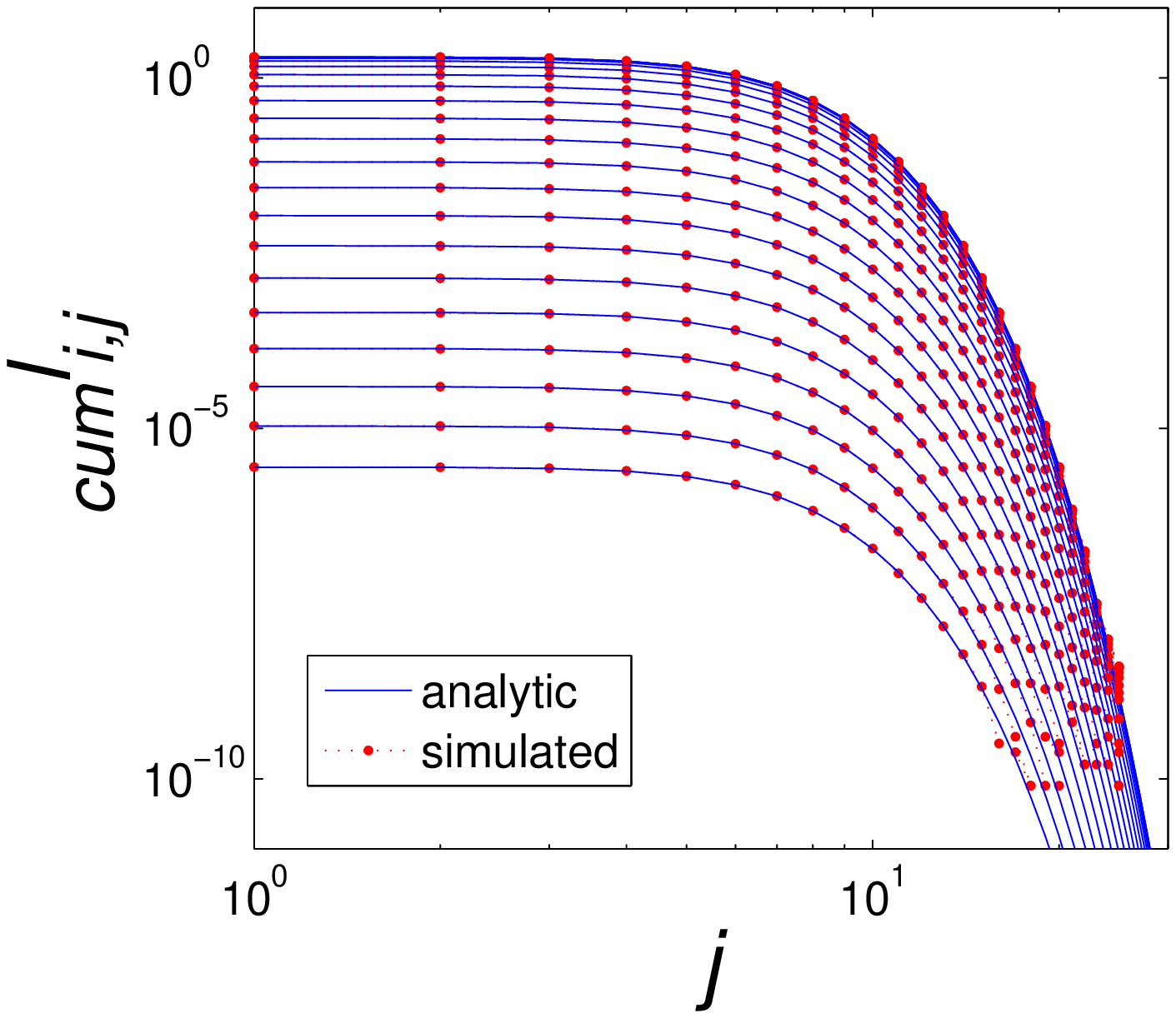}
\caption{ \label{fig:ER} Left: the analytically-derived, normalised link-space matrix for the Erd\H{o}s and R\'enyi (ER) random graph with mean degree equal to $5$. This could be filled to arbitrary size but is here truncated to maximum degree of $30$. Right: comparison of the cumulative link-space matrices for the analytic solution and a simulation of the algorithm. The simulation comprises an ensemble average of $500$ networks each consisting of $10^7$ nodes. The probability that a link exists between any pair of nodes is $\alpha = 5\times10^{-7}$ and the maximum degree obtained was $25$. The first $20$ rows ($i$ values) of the cumulative link-space are illustrated and, again, the finite nature of the simulation is apparent for high $j$.}
\end{center} 
\end{figure}

Using the random attachment probability kernel $\Theta_i(t) = X_i(t)/N(t)$, the node-space master equation for the number of nodes of degree $i$ can be written
\begin{eqnarray}\label{eqn:lsacc1}
\langle X_i(t+1)\rangle&=&X_i(t)+\frac{\langle m(t) \rangle X_{i-1}(t)}{N(t)} -\frac{\langle m(t) \rangle X_{i}(t)}{N(t)}\nonumber \\
{}&{}& +P\{m(t)=i\}.
\end{eqnarray}
The time-dependent solution of the degree distribution for this model is simply 
\begin{eqnarray}
c_i(t)&=& \frac{e^{-\alpha t}(\alpha t)^i}{i!},
\end{eqnarray}
which is what we would expect for the random graph with mean degree of $\alpha t$ in the large size limit~\cite{Boll2}. 
We can similarly write the link-space master equation for this model:
\begin{eqnarray}\label{eqn:lsacc}
\langle L_{i,j}(t+1)\rangle &=&L_{i,j}(t) +\frac{\langle m(t) \rangle \Theta_{i-1}(t) L_{i-1,j}}{X_{i-1}}\nonumber \\
{}&{}& +\frac{\langle m(t) \rangle \Theta_{j-1}(t) L_{i,j-1}}{X_{j-1}} -\frac{\langle m(t) \rangle \Theta_{i}(t) L_{i,j}}{X_{i}} \nonumber\\
{}&{}& -\frac{\langle m(t) \rangle \Theta_{j}(t) L_{i,j}}{X_{j}} + i \Theta_{j-1}P\{m(t)= i\}\nonumber \\
{}&{}&+j \Theta_{i-1}P\{m(t)= j\}.
\end{eqnarray}
The last two terms reflect the expected number of new $i\leftrightarrow j$ links being formed between the new node and existing nodes. Recalling that the normalised link-space matrix is given by $l_{i,j}(t) = L_{i,j}(t)/M(t)$ where the number of links, $M(t)$, at time $t$ can be approximated as $M(t)\approx \alpha t^2/2$ and making use of the previously derived, time-dependent degree distribution such that $P\{m(t)=j\}=c_j(t)$, (\ref{eqn:lsacc}) can be rewritten as
\begin{eqnarray}\label{eqn:lsacc2}
\frac{\textrm{d}}{\textrm{d}t}\left(\frac{\alpha t^2l_{i,j}(t)}{2}\right)&=&\frac{(\alpha t)^2}{2}\big(l_{i-1,j}(t) +l_{i,j-1}(t) -2l_{i,j}(t) \big)\nonumber \\
{}&{}&+\frac{2i j c_i(t)c_j(t)}{\alpha t}.
\end{eqnarray}
The solution to (\ref{eqn:lsacc2}) is found by means of ansatz (the choice of which is explained in Section~\ref{sec:degrees}) and is
\begin{eqnarray}\label{ERLS}
l_{i,j}(t)&=&\frac{2e^{-2\alpha t}(\alpha t)^{i+j-2}}{(i-1)!(j-1)!}.
\end{eqnarray}
This is the normalised link-space matrix for a classical random graph with a mean degree of $\alpha t$ in the large size limit and this is illustrated in Fig~\ref{fig:ER} where a comparison to a simulation is also made. 

%\newpage
\section{Degree correlations and assortativity}\label{sec:degrees}
As discussed in Section~\ref{sec:correlations}, calculation of the mean degree of nearest neighbours of a node as a function of the degree of that node, $\langle k_{nn}\rangle_i$ presents an indication as to the degree correlations of the network. This calculation is straightforward if one has the link-space matrix for a particular network,
\begin{eqnarray}\label{eqn:karnn}
\langle k_{nn}\rangle_i&=&\frac{\sum_j j L_{i,j}}{\sum_j L_{i,j}}\nonumber\\
{}&=&\frac{\sum_j j l_{i,j}}{\sum_j l_{i,j}}.
\end{eqnarray}
\begin{figure}
\begin{center}
\includegraphics[width=0.7\textwidth]{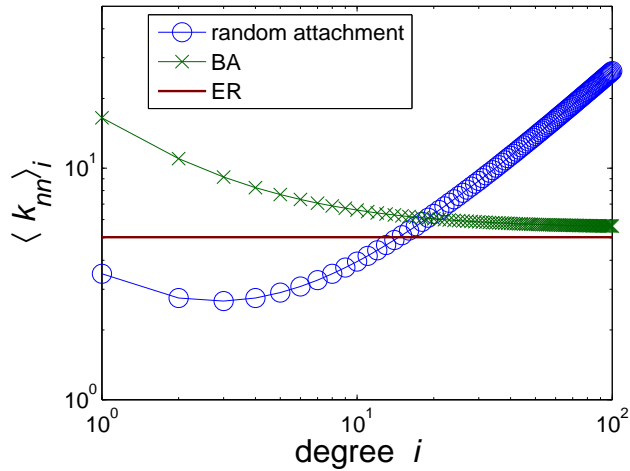}
\caption{ \label{fig:assort2} The mean degree of the nearest neighbours $\langle k_{nn}\rangle_i$ as a function of $i$ for both the preferential and random attachment algorithms and the calssical random graph described in Section~\ref{subsection:nonassort}.}
\end{center}
\end{figure}
This is illustrated in Figure~\ref{fig:assort2}. If the average nearest neighbour degree  $\langle k_{nn}\rangle_i$ is constant, non-assortativity is implied. Certainly, the preferential attachment curve in Figure~\ref{fig:assort2} appears to asymptote to a constant value (as was noted in ~\cite{EK}), whilst that of random attachment continues to increase, implying positively assortative mixing.

In search of a single number representation, Newman~\cite{Newman} further streamlined the measure by simply considering the correlation coefficient of the degrees of the nodes at each end of the links in a given network, although he subtracted one from each first to represent the remaining degree. 	Whilst this might be of interest in some situations, the correlation of the actual degrees is arguably a more general measure although the two do not differ much in practice. Using a normalised Pearson's~\cite{Newman} correlation coefficient, the measure of assortativity is expressed in terms of the degrees ($i$ and $j$) of the nodes at each end of a link, and the averaging is performed over all links in the network.
\begin{eqnarray}
r&=&\frac{\langle ij\rangle-\langle i\rangle \langle j\rangle}{\langle ij \rangle_{assort}-\langle i\rangle_{assort}\langle j\rangle_{assort}}.
\end{eqnarray}
The normalisation factor represents the correlation of a perfectly assortative network, i.e. one in which nodes of degree $i$ are only connected to nodes of degree $i$. This would be of the same degree distribution as the network under consideration and has the same number of nodes and links. The physical interpretation is less than obvious especially as we are considering trees. However, for this assortative network, $l_{i,j} =i c_i\delta(i-j)$, i.e. all off-diagonal elements are zero. Whilst this assortativity measure can be expressed as summations over all links if calculated for a specific realisation of a network, we can express it in terms of the previously derived link-space correlation matrix, namely
\begin{eqnarray}
r&=&\frac{\sum_{i=1}^{\infty}\sum_{j=1}^{\infty}\frac{i j l_{i,j}}{2} - \Big(\sum_{j=1}^{\infty}\frac{j^2 c_j}{2}  \Big)^2      }{\sum_{j=1}^{\infty}\frac{j^3 c_j}{2} - \Big(\sum_{j=1}^{\infty}\frac{j^2 c_j}{2}  \Big)^2}.
\end{eqnarray}
It is interesting that the normalisation factor is degree-distribution specific.
Evaluating this measure numerically, we obtain for the preferential attachment algorithm the value $r_{pa} \approx 0$ and for the random attachment 
algorithm, $r_{ra} \approx 0.6$. So according to this measure of degree mixing, the randomly grown graph is highly assortative and the preferential attachment graph is not.

\begin{figure}
\begin{center}
\includegraphics[width=0.45\textwidth]{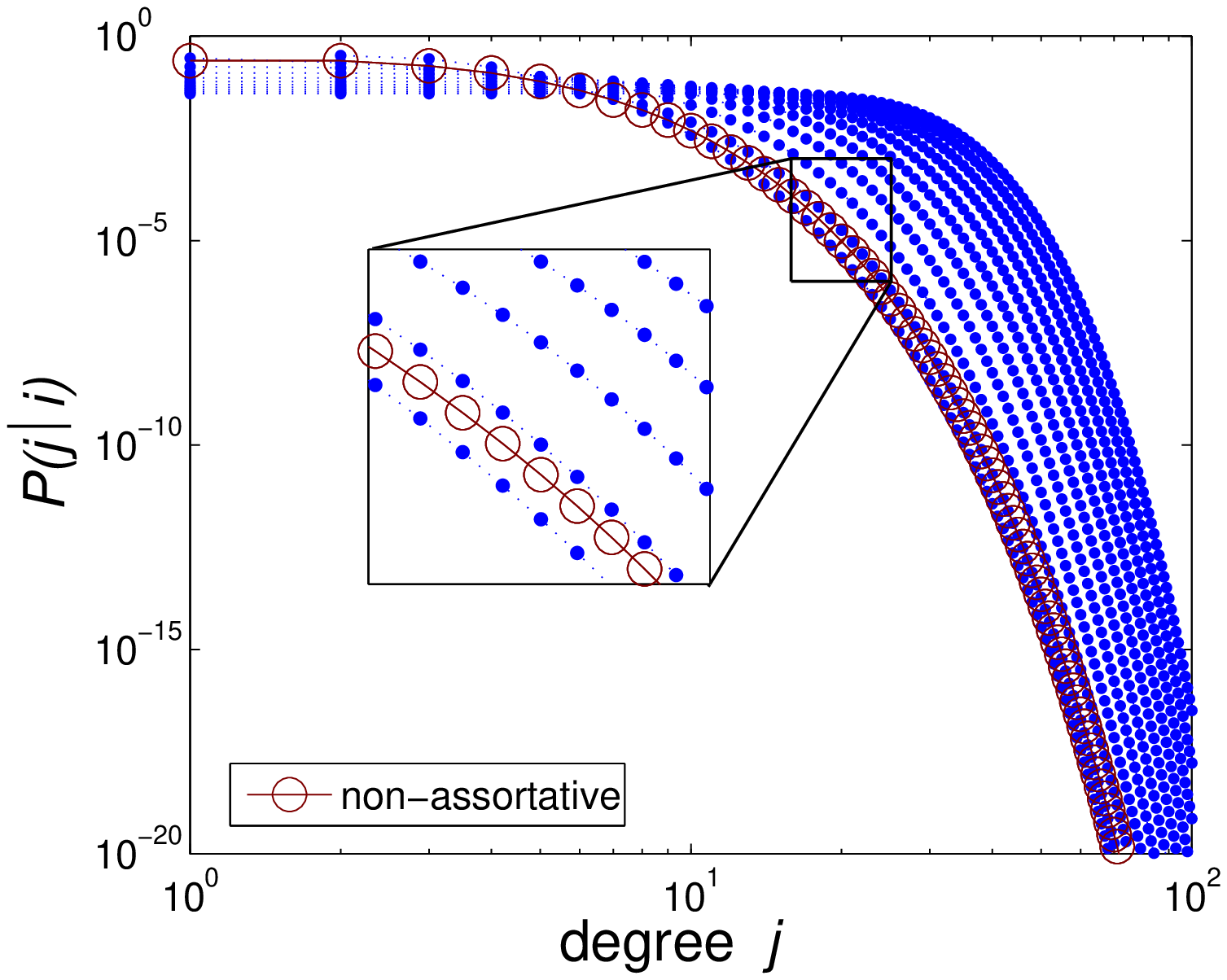}
\includegraphics[width=0.45\textwidth]{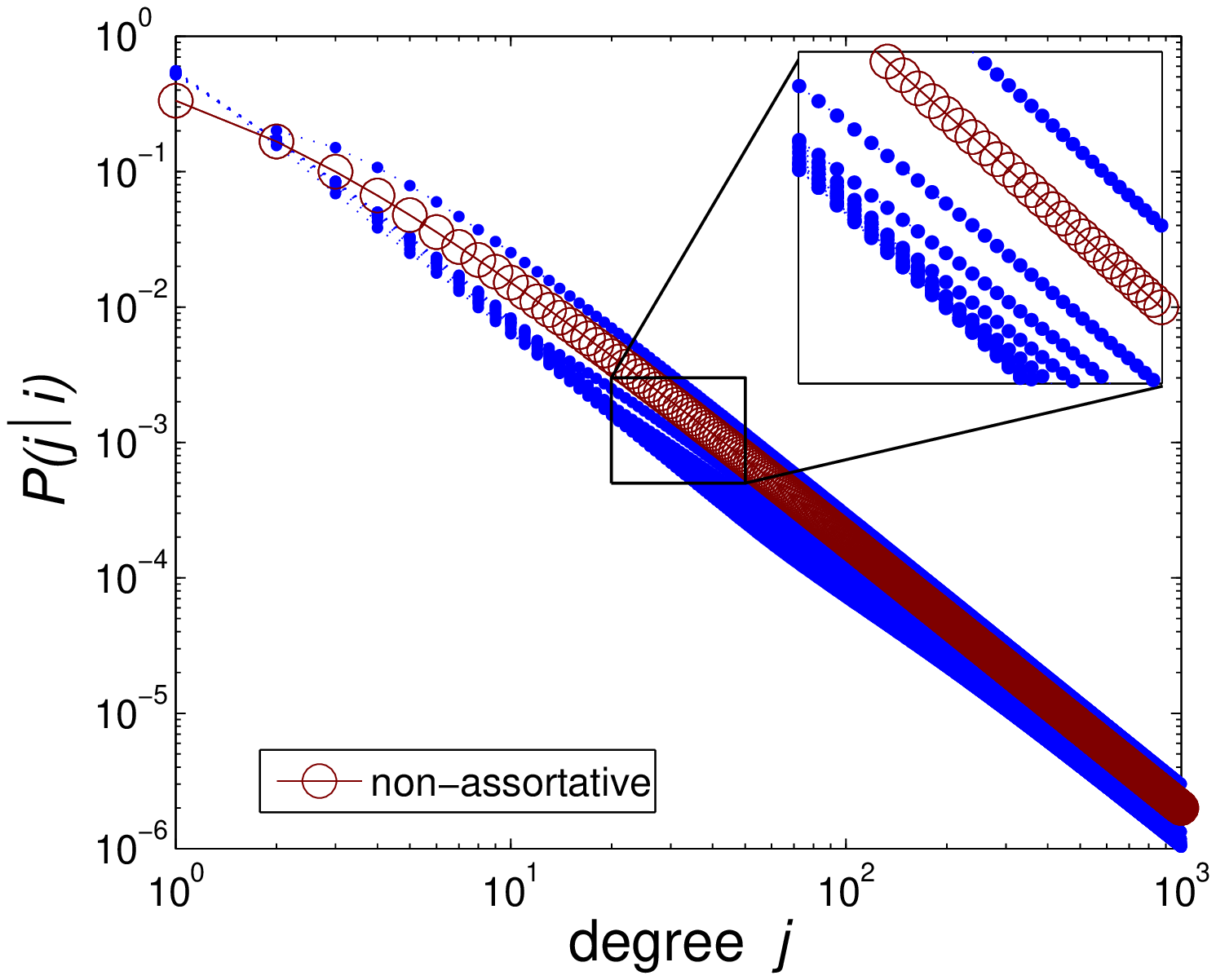}
\caption{ \label{fig:assort1} The conditional probability $P(j\mid i)$ that one end of a randomly selected link has degree $j$ given that the other end is of degree $i$ over the  range $i=\{1,6,11,16,...,51\}$. The left plot is for the random attachment algorithm and the right is preferential attachment. The criterion of non-assortativity of (\ref{eqn:assort1}) is denoted with red circles.}
\end{center}
\end{figure}

The link-space formalism allows us to address two-vertex correlations in a more powerful way in that we can calculate the conditional, two-point vertex degree distribution, $P(j|i)$ which has been traditionally difficult to measure~\cite{boguna:2003}. We can write this joint probability, i.e. the probability that a randomly chosen edge is connected to a node of degree $j$ given that the other end is connected to a node of degree $i$, in terms of the link-space matrix (normalised or not) as $P(j|i)= L_{i,j} / (i X_i) =  L_{i,j}/ \sum_{j=1}^{\infty} L_{i,j} =  l_{i,j}/ \sum_{j=1}^{\infty}  l_{i,j} $. For total number of edges aproximately equal to the total number of nodes, this can be approximated as $l_{i,j} / (i c_i)$. This is illustrated in Fig.~\ref{fig:assort1} for the scenarios of random and preferential attachment.

Consider selecting an edge at random in the network. If we select one end of the edge at random, the probability that this node has degree $j$ will be proportional to $j$ since higher degree nodes have, by definition, more links connected to them than low degree nodes. Consider now only a subset of all edges with one end attached to a node of degree $i$. If there were no correlations present, the probability that the other end is attached a node of degree $j$ is again proportional to $j$. This is the criterion of non-assortativity \cite{Vazquez2} and for total number of edges $\approx$ total number of nodes can be expressed
\begin{eqnarray}
\label{eqn:assort1}
P(j\mid i)&=& P(j) \nonumber\\
{}&=& \frac{j c_j}{2}.
\end{eqnarray}
This is also the criterion that a one step random walk replicates preferential attachment. 
%For a given network, we can write the conditional probability $P(j|i)$, i.e. the probability that a randomly chosen edge is connected to a node of degree $j$ given that the other end is connected to a node of degree $i$, in terms of the link-space matrix 
%\begin{eqnarray}
%\label{eqn:assort2}
%P(j\mid i)&=& \frac{L_{i,j}}{i X_i}\nonumber\\
%{}& =&\frac{l_{i,j}}{i c_i},
%\end{eqnarray}
%where the latter expression holds for total number of edges $\approx$ total number of nodes.
%This criterion is illustrated in Figure~\ref{fig:assort1} for the pedagogical scenarios of random and preferential attachment and confirms that neither is perfectly non-assortative.

\subsection{Perfect non-assortativity\label{subsection:nonassort}}
It is interesting to ask whether a perfectly non-assortative network can be generated. Assuming that such a network will not have equal numbers of nodes and links, we must rewrite the conditions of non-assortativity accordingly for total number of edges $M$: 
\begin{equation}
\label{eqn:assort3}
P(j|i)=P(j)=\frac{j X_j}{2 M}= \frac{j N c_j}{2 M}.
\end{equation}
Recalling our definition of the normalised link-space matrix that $l_{i,j} = L_{i,j}/M$, we can express this conditional probability $P(j|i)$ in terms of this matrix as
\begin{equation}\label{eqn:assort4}
P(j|i)= \frac{L_{i,j}}{i X_i}=\frac{M l_{i,j}}{iNc_i}.
\end{equation}
As such, we can now write for $l_{i,j}$
\begin{eqnarray}
\label{eqn:assort5}
l_{i,j}&=& \bigg(\frac{N}{M}\bigg)^2 \frac{i c_i j c_j}{2}.
\end{eqnarray}
That is, for \emph{any} degree distribution, a normalised link-space matrix can be found which is representative of a perfectly non-assortative network \footnote{A network can be constructed for an arbitrary, normalised link-space matrix. A diagonal element $l_{i,i}$ can be simply obtained through the addition of fully connected components of $i+1$ nodes. Similarly an off-diagonal element $l_{i,j}$ can be generated through suitable rewiring of $i$-degree and $j$-degree cliques.}. This might provide an alternative to the network randomisation technique of Maslov and Sneppen to provide a ``null model network" \cite{maslov:2002}.  The expression of (\ref{eqn:assort5}) provides the ansatz solution for the normalised link-space matrix to the master equations for the ER random graph in Section~\ref{subsec:ER}.
%\newpage
\section{Decaying networks}\label{sec:decay}
Although somewhat counter-intuitive, it is possible to find steady states of networks whereby nodes and/or links are removed from the system. Aside from the obvious situation of having no nodes or edges left, we would like to investigate the possible existence of a network configuration whose link-space matrix and, subsequently, degree distribution are static with respect to the decay process. The concept that decay processes are highly influential on a network's structure has been considered before although this has typically only been investigated in conjunction with simultaneous growth~\cite{Duplication, Laird, May}. Here we shall employ the link-space formalism to examine the effect of some simple, decay-only scenarios, specifically the two simplest cases -- random link removal and random node removal.
\subsection{Random Link Removal (RLR)}\label{subsec:rlr}
\begin{figure}
\begin{center}
\includegraphics[width=0.7\textwidth]{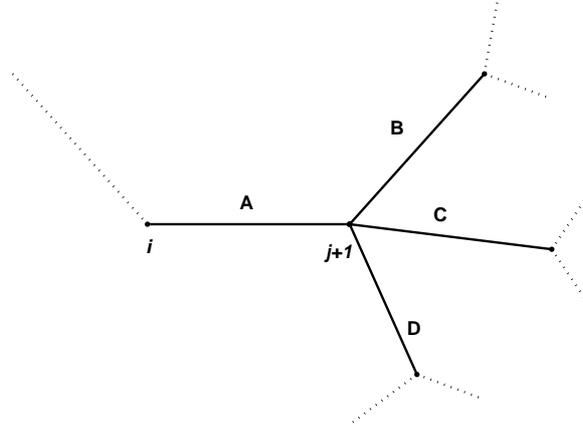}
\caption{ \label{fig:decay1} Removal of any of the links $B,C,D$ will result in the link $A$ becoming an $i\leftrightarrow j$ link.}
\end{center}
\end{figure}

Consider an arbitrary network. We select $w$ links at random and remove them. We shall implement the link-space to analyse the evolution of the network. Consider the link-space element $L_{i,j}(t)$ denoting the number of links from nodes of degree $i$ to nodes of degree $j$. Clearly this can be decreased if an $i\leftrightarrow j$ link is removed. Also, if a  $k\leftrightarrow i$ link is removed and that $i$ node has further links to $j$ degree nodes, then those that were $i\leftrightarrow j$ links will now be $i-1\leftrightarrow j$, similarly for $k\leftrightarrow j$ links being removed. However, if the link removed is a $k\leftrightarrow j+1$ link and that $j+1$ degree node is connected to a $i$ degree node, then when the $j+1$ node becomes an $j$ node, that link will become an $i\leftrightarrow j$ link increasing $L_{i,j}$ as shown in Figure~\ref{fig:decay1}. Let us assume that we are removing links at random from the network comprising $N(t)$ vertices and $M(t)$ links. A non-random link selection process could be incorporated into the master equations using a probability kernel. The master equation for this process can be written in terms of the expected increasing and decreasing contributions:
\begin{eqnarray}\label{eqn:RLR1}
\langle L_{i,j} (t+1)\rangle&=& L_{i,j}(t) + w\sum_{k}\frac{L_{k,j+1}(t) j L_{i,j+1}(t)}{M(t) (j+1) X_{j+1}(t)}\nonumber\\
{}&{}&+w\sum_{k}\frac{L_{k,i+1}(t) i L_{i+1,j}(t)}{M(t) (i+1) X_{i+1}(t)}\nonumber\\
{}&{}&-w\sum_{k}\frac{L_{k,j}(t) (j-1) L_{i,j}(t)}{M(t) (j) X_{j}(t)}\nonumber\\
{}&{}&-w\sum_{k}\frac{L_{k,i}(t) (i-1) L_{i,j}(t)}{M(t) (i) X_{i}(t)}\nonumber\\
{}&{}&-\frac{w L_{i,j}(t)}{M(t)}.
\end{eqnarray}
This simplifies to
\begin{eqnarray}\label{eqn:RLR2}
\langle L_{i,j}(t+1)\rangle &=& L_{i,j}(t) + \frac{w j L_{i,j+1}(t)}{M(t)}\nonumber\\ 
{}&{}&+\frac{w i L_{i+1,j}(t)}{M(t)}-\frac{w (i-1) L_{i,j}(t)}{M(t)}\nonumber\\
 {}&{}& -\frac{w (j-1) L_{i,j}(t)}{M(t)}-\frac{wL_{i,j}(t)}{M(t)}.
\end{eqnarray}
This expression can be simply understood by considering Figure~\ref{fig:decay1}. For each $i\leftrightarrow j+1$ link (labelled $A$ in the figure), there are $j$ links ($B,C,D$) whose removal would make link $A$ an $i\leftrightarrow j$ link. This would be the same for all of the  $i\leftrightarrow j+1$ links, of which there are $L_{i,j+1}(t)$ in the network. 
The expected increase in $L_{i,j}$ through this process corresponds to the second term on the right hand side of (\ref{eqn:RLR2}). The other terms are equally simply obtained, with the last referring to the physical removal of an $i\leftrightarrow j$ link.

\begin{figure}
\begin{center}
\includegraphics[width=0.65\textwidth]{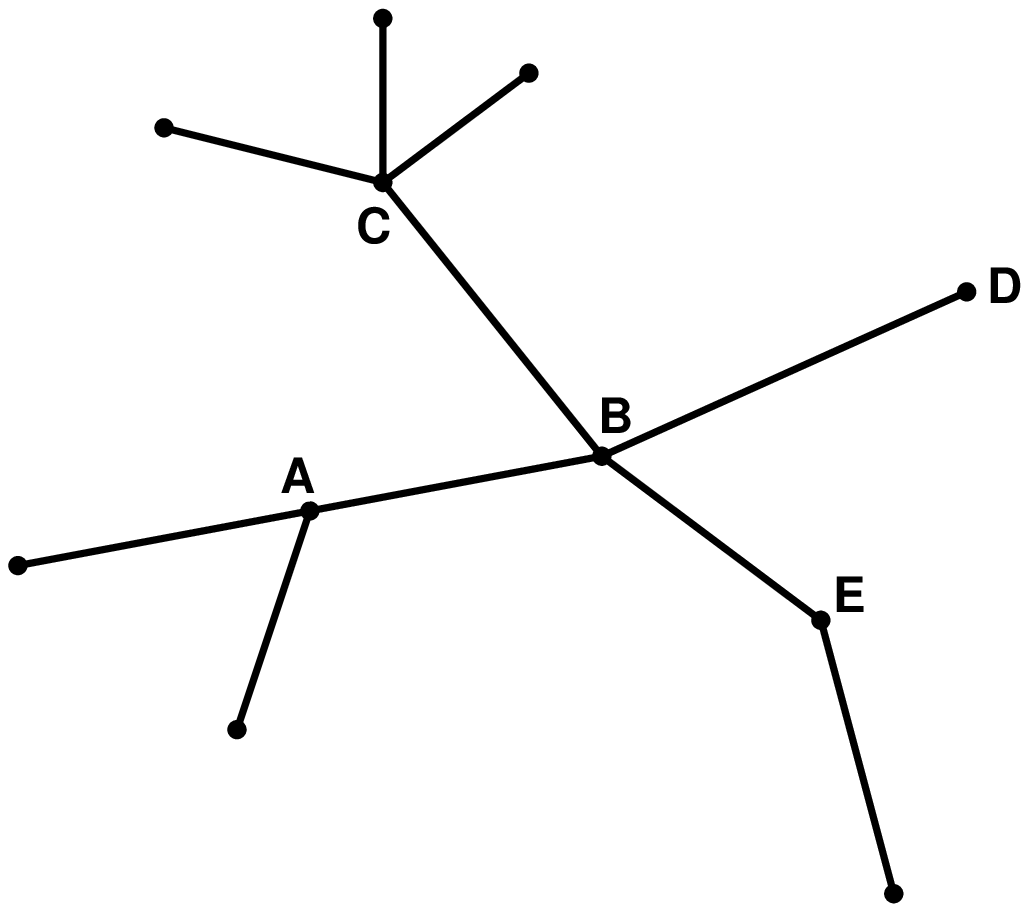}
\caption{ \label{fig:simplenet4} A simple network which we shall examine to check the validity of the master equation (\ref{eqn:RLR2}) for the evolution of the link-space matrix for the random link removal process.}
\end{center} 
\end{figure}

To check that all is well, we can explicitly look at the evolution of the link-space matrix for the simple example network discussed in Section~\ref{sec:Saramaki} and shown again in Figure~\ref{fig:simplenet4}.
Prior to removing any links, the link-space can be simply written as in (\ref{eqn:simpleL}).
 \begin{eqnarray}\label{eqn:simpleL}
 \mathbf{L}(t)&=& \left( \begin{array}{cccc}
 0&1 & 2 &4\\
 1&0 & 0 &1\\
 2&0 & 0 &1\\
 4&1 & 1 &2
 \end{array} \right).
 \end{eqnarray}
The expected link-space matrix after removing a link at random can be interpreted as the ensemble average of all possibilities. By removing a single link from the original network, the resulting link-space matrix for the new network can be easily generated. Replacing that link and performing this process for all links and simply averaging all the resulting matrices, we can generate (albeit somewhat laboriously) the desired expected link-space matrix for the network at the next timestep. This yields the result of (\ref{eqn:simpleLt1}) which represents the average over all possible resulting networks after one link has been removed at random. This concurs with the master equation for this process given in (\ref{eqn:RLR2}).
\begin{eqnarray}\label{eqn:simpleLt1}
 \langle \mathbf{L} (t+1)\rangle &=& \frac{1}{10}\left( \begin{array}{cccc}
 2&12 & 26 &25\\
 12&0 & 3 &7\\
 26&3 & 6 &10\\
 25&7 & 10 &6
 \end{array} \right).
 \end{eqnarray}

\begin{figure}
\begin{center}
\includegraphics[width=0.7\textwidth]{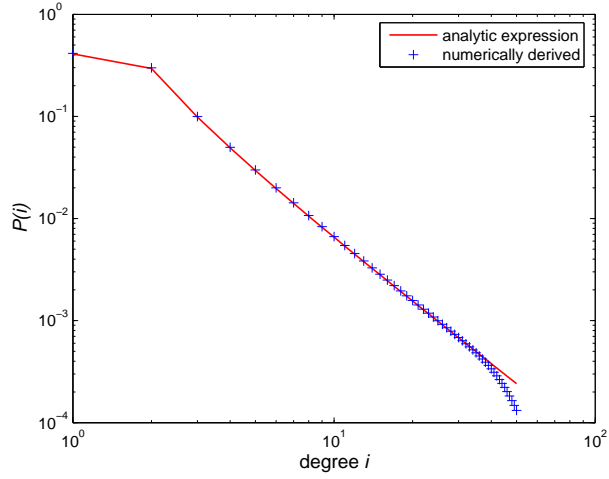}
\caption{ \label{fig:rlrdegs} Comparison of the degree distributions from the numerically populated link-space matrix and the analytic results of (\ref{eqn:analrlr}) for the Random Link Removal model. The two node components have been neglected in both the numerical matrix population and analytically derived expression.}
\end{center}
\end{figure}

To investigate the possibility of a steady state solution, we make a similar argument as before for growing networks but this time for the process of removing $w$ links per timestep.
\begin{eqnarray}
L_{i,j}(t)&=&l_{i,j}M(t),\nonumber\\
{}&=&l_{i,j}(M_0-wt),\nonumber\\
\frac{dL_{i,j}}{dt}&=&-wl_{i,j},\nonumber\\
{}&\approx & L_{i,j}(t+1)-L_{i,j}(t).
\end{eqnarray}
The expression of (\ref{eqn:RLR2}) can be reduced to
\begin{eqnarray}\label{eqn:lsrlr}
l_{i,j}&=&\frac{i l_{i+1,j}+j l_{i,j+1}}{i+j-2}.
\end{eqnarray}
The number of $1\leftrightarrow 1$ links does not reach a steady state. This might be expected, as the removal process for such links requires them to be physically removed as opposed to the process by which the degree of the node at one end of the link is reduced. Subsequently, the value $l_{1,1}$ would increase in time. However, we can investigate the properties of the rest of the links in the network which do reach a steady state by neglecting these two node components. In a similar manner to ~\cite{Krapivsky} we can make use of a substitution to find a solution to this recurrence equation, namely for $i+j\ne2$.
\begin{eqnarray}\label{eqn:landm}
l_{i,j}&=& q_{i,j}\frac{(i+j-3)!}{(i-1)!(j-1)!},
\end{eqnarray}
and we obtain
\begin{eqnarray}
q_{i,j}&=&q_{i+1,j}+q_{i,j+1}.
\end{eqnarray}
This has a trivial solution
\begin{eqnarray}\label{eqn:msolution}
q_{i,j}&=&\frac{A}{2^{i+j}}.
\end{eqnarray}
The link-space for this system can then be written for $i+j\ne2$
\begin{eqnarray}\label{eqn:rlrfull}
l_{i,j}&=&\frac{A}{2^{i+j}}\frac{(i+j-3)!}{(i-1)!(j-1)!}.
\end{eqnarray}
\begin{figure}
\begin{center}
\includegraphics[width=0.7\textwidth]{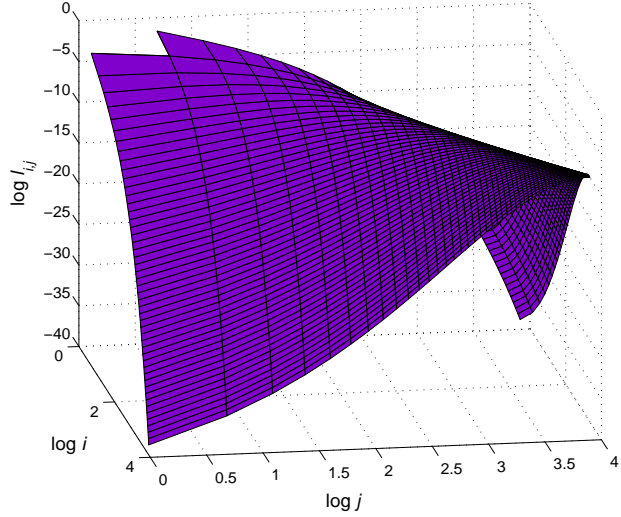}
\caption{ \label{fig:rlr} The shape of the steady state link-space for random link removal from an infinite network numerically populated using (\ref{eqn:rlrfull}) for $i+j\ne2$.}
\end{center}
\end{figure}
The summation over this link-space matrix does not converge, i.e.  $\sum_i \sum_j l_{i,j} \rightarrow \infty$. However, we can still derive the degree distribution and normalise this such that $\sum_i c_i =1$. We can use the form of (\ref{eqn:rlrfull}) to infer the shape of the degree distribution. Recalling that
\begin{eqnarray}\label{eqn:ci3}
c_i&=& \frac{M}{N}\frac{\sum_k l_{i,k}}{i},
\end{eqnarray}
we can write for the degree distribution, neglecting the two node components in the network,
\begin{eqnarray}\label{eqn:rlrdegdist1}
c_1&=&\frac{M A}{N}\sum_{k=2}^{\infty}\frac{(k-2)!}{(k-1)! 2^{1+k}},
%{}&=&\frac{M A}{4N}log(2)\nonumber\\
\end{eqnarray}
\begin{eqnarray}\label{eqn:rlrdegdist2}
c_i&=&\frac{M A}{Ni}\sum_{k=1}^{\infty}\frac{(i+k-3)!}{(i-1)!(k-1)! 2^{i+k}} ~~~~~~~~i>1.
%{}&=&\frac{M A}{4N}\frac{1}{i(i-1)}\nonumber
\end{eqnarray}
The equation (\ref{eqn:rlrdegdist1}) can be easily solved by considering the Maclaurin Series (Taylor Series about zero) of the function $\log(1+x)$ (as studied by Mercator as early as 1668) and evaluating for $x=-\frac{1}{2}$,
\begin{eqnarray}
\log(1+x)&=&\sum_{k'=1}^{\infty}\frac{(-1)^{k'}}{k'} x^{k'}\nonumber\\
\log\bigg(\frac{1}{2}\bigg)&=& -\sum_{k'=1}^{\infty}\frac{1}{k'~2^{k'}}\nonumber\\
{}&=&-\log(2).
\end{eqnarray}
Rearranging (\ref{eqn:rlrdegdist1}) and letting $k=k'+1$,
\begin{eqnarray}
c_1&=&\frac{M A}{N}\sum_{k=2}^{\infty}\frac{1}{(k-1) 2^{1+k}}\nonumber\\
{}&=&\frac{M A}{N}\sum_{k'=1}^{\infty}\frac{1}{k'2^{k'+2}}\nonumber\\
{}&=&\frac{M A}{4N}\log(2).
\end{eqnarray}
For nodes with degree greater than one, the solution of (\ref{eqn:rlrdegdist2}) requires a simple proof by induction. Consider the function $Q(i')$ defined for positive integer $i'$ as
\begin{eqnarray}
Q(i')&=& \sum_{k'=0}^{\infty} \frac{{i'+k' \choose k'}}{2^{k'+i'}},
\end{eqnarray}
where ${i'+k' \choose k'}$ denotes the conventional binomial coefficients. We can write $Q(i'+1)$ with similar ease:
\begin{eqnarray}
Q(i'+1)&=& \sum_{k'=0}^{\infty} \frac{{i'+k'+1 \choose k'}}{2^{k'+i'+1}}.
\end{eqnarray}
As such,
\begin{eqnarray}
2Q(i'+1)-Q(i')&=& \sum_{k'=0}^{\infty}\bigg({i'+k'+1 \choose k'}-{i'+k'\choose k'} \bigg)\frac{1}{2^{k'+i'}}\nonumber\\
{}&=& \sum_{k'=1}^{\infty}~{i'+k'\choose k'-1} \frac{1}{2^{k'+i'}}.
\end{eqnarray}
A quick substitution of $k''=k'-1$ leads to
\begin{eqnarray}
2Q(i'+1)-Q(i')&=&\sum_{k''=0}^{\infty}~{i'+k''+1 \choose k''} \frac{1}{2^{k''+i'+1}}\nonumber\\
{}&=&Q(i'+1)\nonumber\\
\Rightarrow~~Q(i'+1)&=&Q(i'),\nonumber\\
Q(1)&=&2\nonumber\\
{}&=&Q(i') ~~~\forall~i' . 
\end{eqnarray}
Rearranging (\ref{eqn:rlrdegdist2}) and using some simple substitutions, $i'=i-2$ and $k'=k-1$ we can derive the following
\begin{eqnarray}
c_i&=&\frac{M A}{Ni}\sum_{k=1}^{\infty}\frac{(i+k-3)!}{(i-1)!(k-1)! 2^{i+k}}\nonumber\\
{}&=&\frac{M A}{Ni(i-1)}\sum_{k=1}^{\infty}\frac{(i+k-3)!}{(i-2)!(k-1)! 2^{i+k}}\nonumber\\
{}&=&\frac{M A}{Ni(i-1)}\sum_{k'=0}^{\infty}\frac{{i'+k' \choose k'}}{ 2^{i'+k'+3}}\nonumber\\
{}&=&\frac{M A}{4N}\frac{1}{i(i-1)}.
\end{eqnarray}
We can normalise this degree distribution for the network without the two node components (the $1\leftrightarrow 1$ links):
\begin{eqnarray}\label{eqn:analrlr}
A&=& \frac{4N}{M (1+\log(2))},\nonumber\\
c_1&=&\frac{\log(2)}{1+\log(2)},\nonumber\\
c_i&=&\frac{1}{(1+\log(2))i(i-1)}~~~~~~~~i>1.
\end{eqnarray}
Although difficult to compare to simulations, we can populate a link-space matrix numerically using (\ref{eqn:rlrfull}) as illustrated in Figure~\ref{fig:rlr}. With this, we can compare the degree distributions using (\ref{eqn:ci3}) and those of (\ref{eqn:analrlr}) as shown in Figure~\ref{fig:rlrdegs}. 

\subsection{Random Node Removal (RNR)}\label{subsec:RNR}
In a similar manner to Section~\ref{subsec:rlr} we will now discuss the possibility of creating such a steady state via a process of removing nodes (and all of their links)  from an existing network at the rate of $w$ nodes per timestep. Clearly, removing some node which has a link to an $(i+1)$-degree node which in turn has a link to a $j$-degree node can increase the number of $i\leftrightarrow j$ links in the system. The other processes which can increase or decrease the number of links from $i$ degree nodes to $j$ degree nodes can be similarly easily explained. We consider that we select a node of some degree $k$ for removal with some probability kernel $\Theta_k$ (as in the growing algorithms of Section ~\ref{sec:linkspace}). The master equation for such a process can thus be written for general node selection kernel:
\begin{eqnarray}\label{eqn:noderemovemaster}
\langle L_{i,j}(t+1) \rangle &=& L_{i,j}(t)~-~w\frac{\Theta_i(t) L_{i,j}(t)}{X_i(t)}-~w\frac{\Theta_j(t) L_{i,j}(t)}{X_j(t)}\nonumber\\
{}&{}&+~w\sum_k \frac{\Theta_k(t)}{X_k(t)} \bigg\{ L_{k,i+1}(t) \frac{i L_{i+1,j}(t)}{(i+1)X_{i+1}(t)}+ L_{k,j+1}(t) \frac{j L_{i,j+1}(t)}{(j+1)X_{j+1}(t)}\bigg\} \nonumber\\
{}&{}&-~w\sum_k \frac{\Theta_k(t)}{X_k(t)} \bigg\{ L_{k,i}(t) \frac{(i-1) L_{i,j}(t)}{iX_i(t)} + L_{k,j}(t) \frac{(j-1)L_{i,j}(t)}{j X_j(t)}\bigg\}.
\end{eqnarray}

We now select a kernel for selecting the node to be removed, namely the random kernel although the approach is general to any node selection procedure. For this process $\Theta_k~=~c_k$ as a node is selected purely at random. The steady state assumptions must be clarified slightly. As before,
\begin{eqnarray}
L_{i,j}(t)&=&l_{i,j}M(t).
\end{eqnarray}
However, because we can remove more than one link (through removing a high degree node for example) we must approximate for $M(t)$. Using the random removal kernel, we can assume that on average, the selected node will have degree $\langle k\rangle(t)=\frac{2M(t)}{N(t)}$ and we can use this to calculate the number of remaining links in the network.
\begin{eqnarray}\label{eqn:rnrsteady}
L_{i,j}(t)&=&l_{i,j}(M_0-w\langle k\rangle(t)) \nonumber\\
{}&=&l_{i,j}(M_0-\frac{2 w M (t)}{N(t)}t),\nonumber\\
\frac{dL_{i,j}}{dt}&=&-\frac{2wM(t)}{N(t)}l_{i,j}\nonumber\\
{}&\approx & L_{i,j}(t+1)-L_{i,j}(t).
\end{eqnarray}

\begin{figure}
\begin{center}
\includegraphics[width=0.7\textwidth]{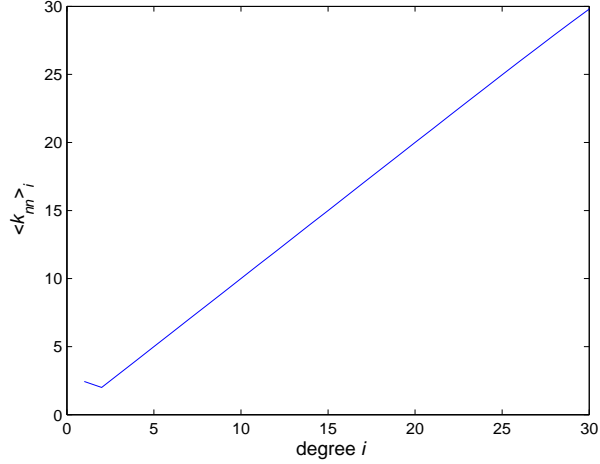}
\caption{ \label{fig:kbarnnrnr} The mean degree of the neighbours of nodes of degree $i$ for the random link and random node removal algorithms.}
\end{center}
\end{figure}

Making use of the link-space identities and (\ref{eqn:rnrsteady}),  the master equation (\ref{eqn:noderemovemaster}) can be written in simple form:
\begin{eqnarray}\label{eqn:rnr}
l_{i,j}&=&\frac{i l_{i+1,j}+ j l_{i,j+1}}{i+j-2}.
\end{eqnarray}
Clearly, this is  identical to (\ref{eqn:lsrlr}) for the random link removal model of Section~\ref{subsec:rlr} and consequently, the analysis of the degree distribution will be the same too. 
By considering the average degree of the neighbours of nodes of degree $i$, denoted $\langle k_{nn}\rangle_i$ and making use of (\ref{eqn:karnn}) in Section~\ref{sec:degrees} we can see that both the random link removal  and the random node removal algorithms generate highly assortative networks as evidenced in Figure~\ref{fig:kbarnnrnr}.

\section{The Growing Network with Redirection model (GNR)}\label{sec:Redner}
In ~\cite{Krapivsky}, Krapivsky and Redner suggested a model of network growth without prior global knowledge using directed links. Within the same paper, they also provide analysis for the evolution of the joint degree distribution of simpler, directed growing networks ($GN$), which is similar to the link-space analysis outlined above.
The \emph{GNR} algorithm is implemented as follows:
\begin{enumerate}
\item Pick a node $\kappa$ within the existing network at random.
\item With probability $a$ make a directed link pointing to that node or
\item Else, pick the ancestor node of $\kappa$  (the node which its out-link is directed to) and make a directed link pointing to that node.
\end{enumerate}

\begin{figure}
\begin{center}
\includegraphics[width=0.65\textwidth]{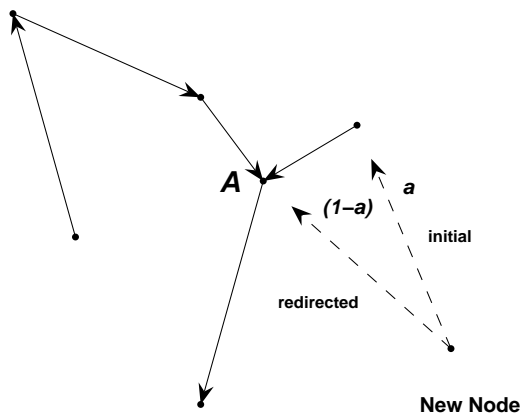}
\caption{ \label{fig:GNR} The Growing Network with Redirection. A random initial node is selected. The new node links to (and points at) this node with probability $a$ or else is redirected to and attaches to the ancestor of the node.}
\end{center}
\end{figure}
We note that because each new node is attached to the existing network with only one link, each node has only one out link. When redirection occurs, the destination is predetermined. The degree distribution for this network is easily obtained. Consider some node of total (in and out) degree $i$ labeled $A$. Its out-degree is $1$ so its in-degree is simply $i-1$. Thus, it must have $i-1$ `daughters'. If any of these $i-1$ daughter nodes is selected initially in the \emph{GNR} algorithm \emph{and} redirection occurs, then the new node will link to and point at $A$. It is straightforward to construct the evolution of the node-space for the total degree of nodes within the system. The attachment probability kernel for the new node linking to any node of degree $i$ within the existing network can be written as
\begin{eqnarray}
\Theta_i &=&  a c_{i} + (1-a)(i-1)c_i.
\end{eqnarray}
The first term on the right hand side corresponds to initially selecting a node of degree $i-1$ and linking to it. The second
term reflects redirection from any of the daughters of any of the degree $i$ nodes. The static equation for the steady state degree distribution becomes
\begin{eqnarray}
c_i&=&  \frac{\big(a+(1-a)(i-2)\big)}{1+a + (1-a)(i-1)}c_{i-1}.
\end{eqnarray}
For $a=0.5$, the familiar recurrence relation of preferential attachment is retrieved and the distribution of total degree of nodes will subsequently be the same too.
%\newpage

\section{The External Nearest Neighbour Model (XNN)}\label{sec:model}
In this section we introduce a model that makes use of only local information about node degrees as microscopic mechanisms requiring global information are often unrealistic for many real-world networks~\cite{Vazquez}. It therefore provides insight into possible alternative microscopic mechanisms for a range of biological and social networks. The link-space formalism allows us to identify the transition point at lower power-law exponent degree distributions switch to higher exponent power-law distributions with respect to the BA preferential attachment model. While similar local algorithms have been proposed in the literature \cite{Evans, Saramaki}, the strength of the approach followed here is the ability to describe the inherent degree-degree correlations \cite{Callaway2, Krapivsky, Newman, NewmanReview}.

It is well known that a mixture of random and preferential attachment in a growth algorithm can produce power-law degree distributions with exponents greater than three, $\gamma \in [3, \infty)$. It has often been assumed that a \emph{one step} random walk replicates linear preferential attachment \cite{Saramaki,Vazquez}. This is \emph{not} true. A one step random walk is in fact more biased towards high degree nodes than preferential attachment (as discussed in Section~\ref{sec:Saramaki}) as can be easily seen by performing the procedure on a simple hub and spoke network. In this case, the probability of arriving at the hub tends to one for increasingly large networks rather than a half as would be appropriate for preferential attachment. We can use this bias to generate networks with degree distributions that are overskewed (lower power-law exponent) than the preferential attachment model. A mixture of this approach with random attachment results in a simple model that can span a wide range of degree distributions. This one-parameter network growth model, in which we simply attach a single node at each timestep, does not require prior knowledge of the existing network structure. 

\begin{figure}
\begin{center}
\includegraphics[width=0.3\textwidth]{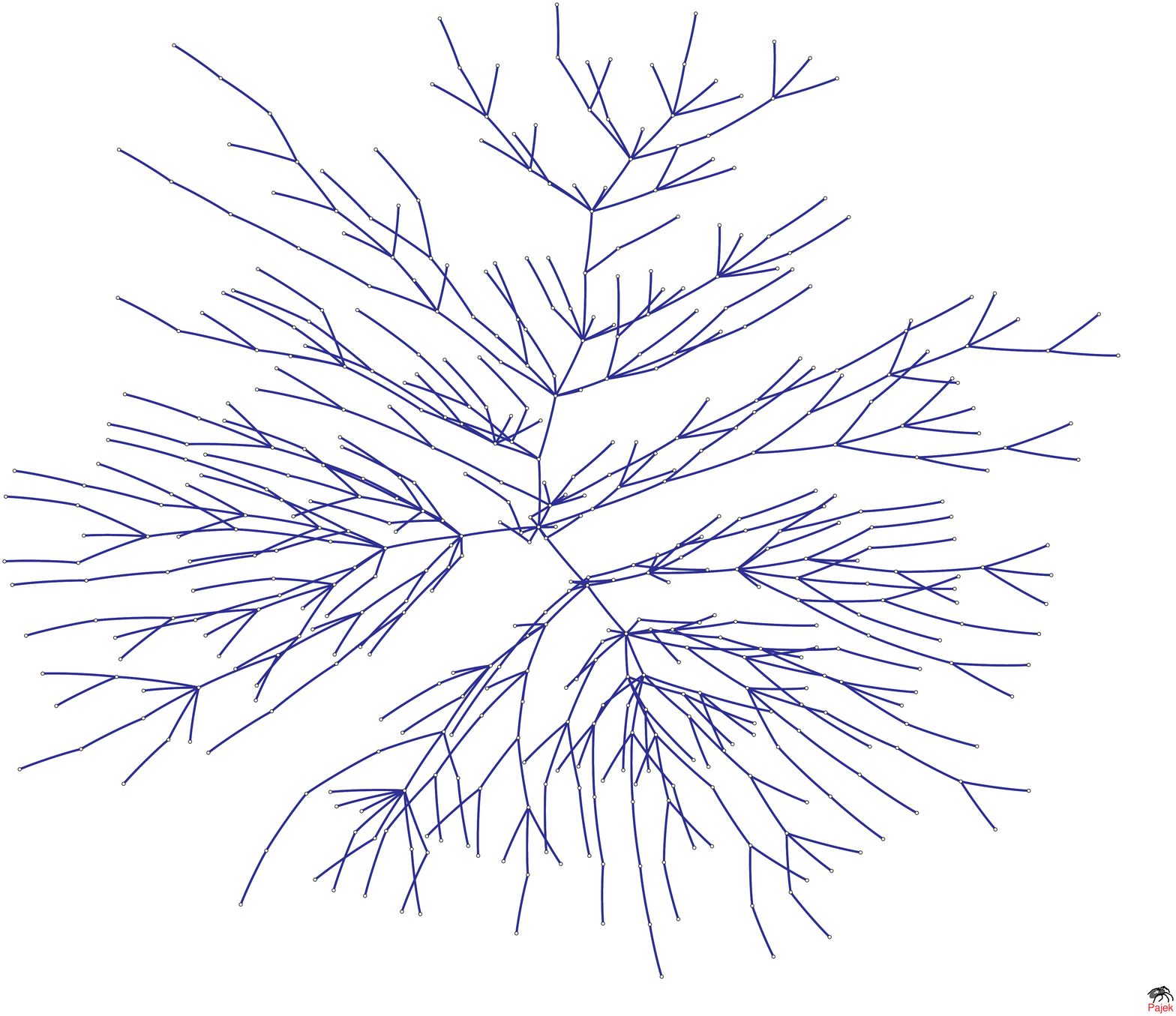}
\includegraphics[width=0.3\textwidth]{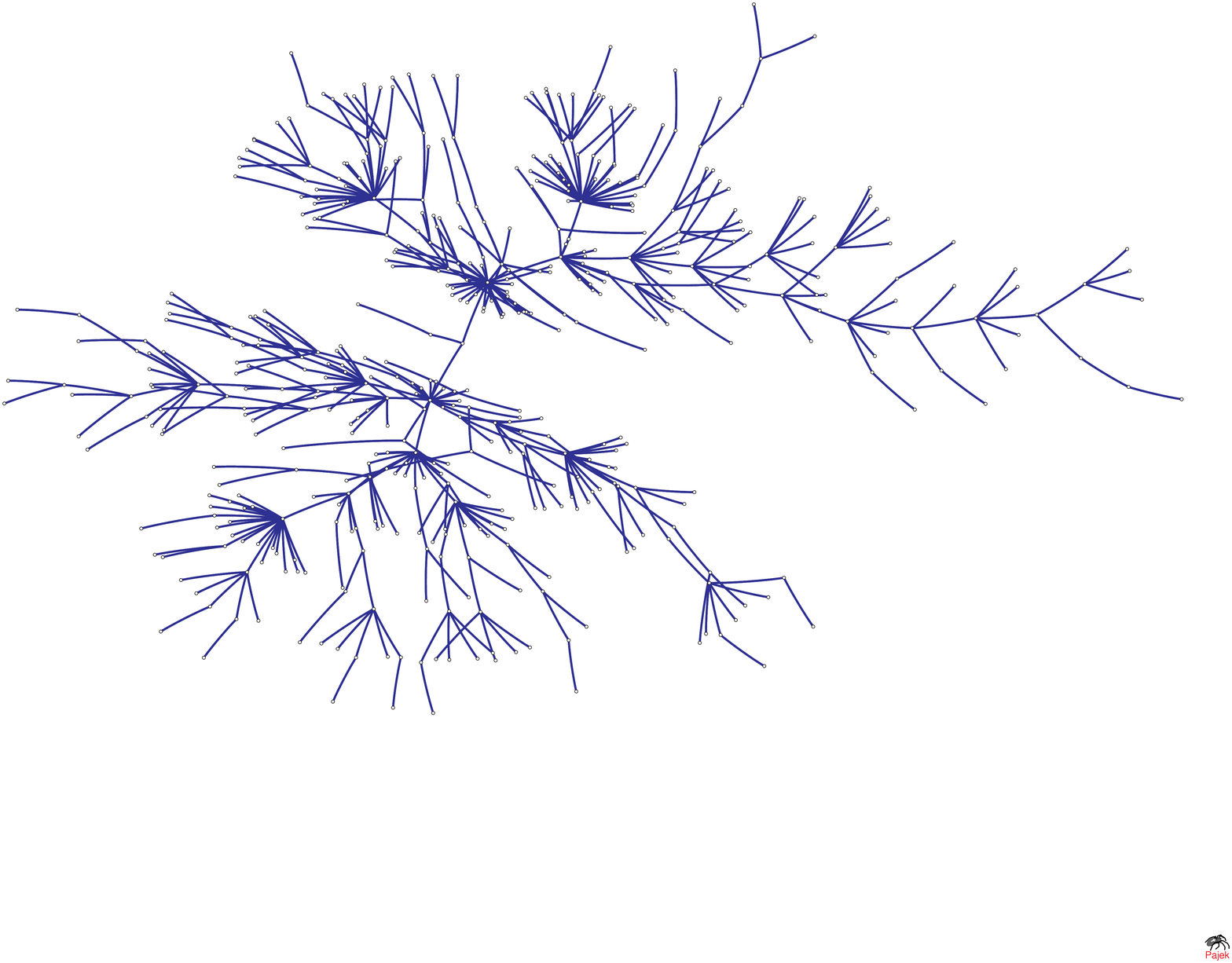}
\includegraphics[width=0.3\textwidth]{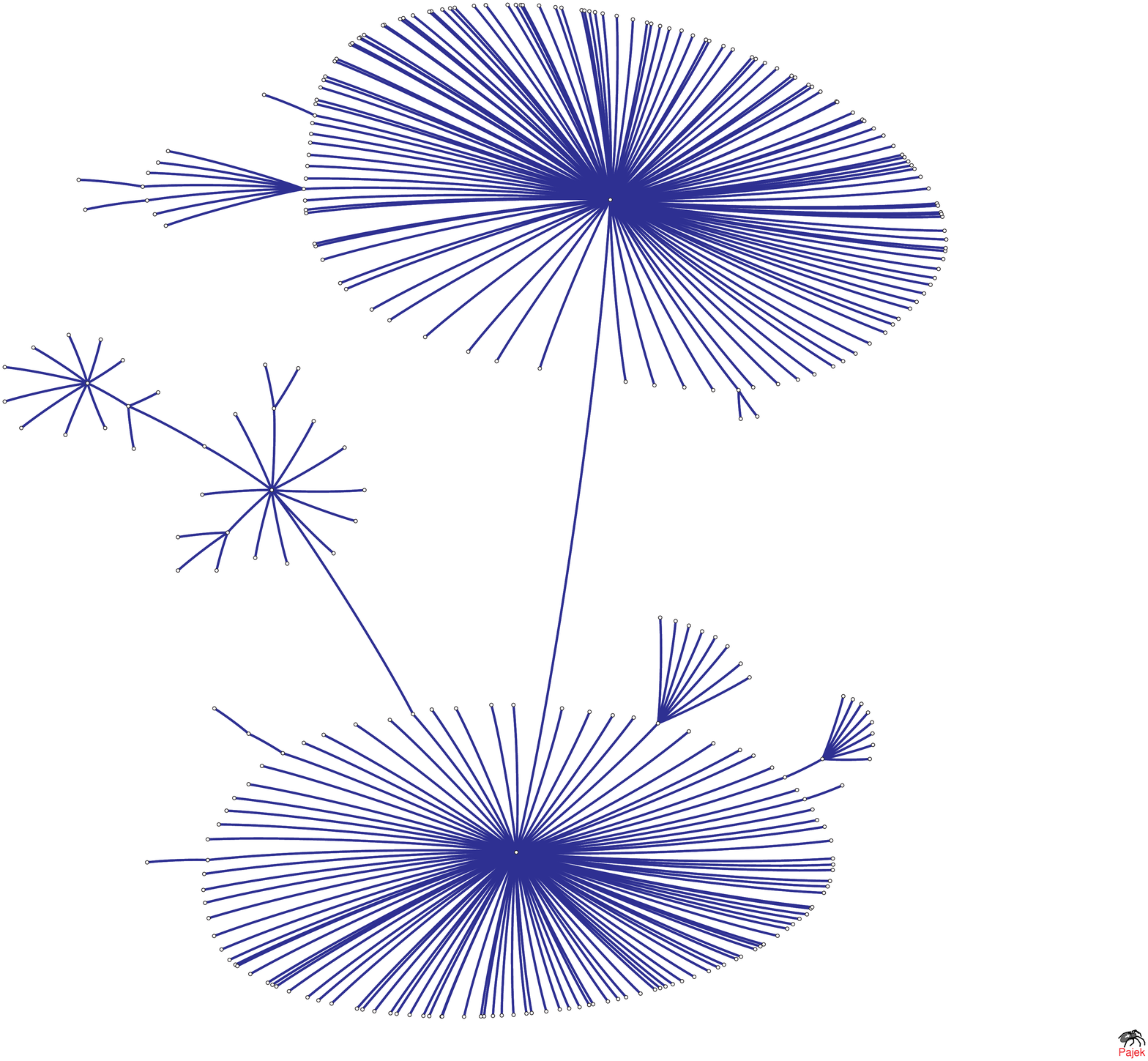}
\caption{ \label{fig:exNN} External Nearest Neighbour: some examples of the graphs generated for 300 nodes. Left is $a =1$ , 
center is $a=0.2$ and right is $a=0$. There is no spatial relevance in these plots which are drawn with pajek~\cite{Batagelj98}. }
\end{center}
\end{figure}

Explicitly, the External Nearest Neighbour algorithm proceeds as follows:
\begin{enumerate}
\item Pick a node $\kappa$ within the existing network at random.
\item With probability $a$ make a link to that node or
\item Else, pick any of the neighbours of $\kappa$ at random and link to that node.
\end{enumerate}
Hence this algorithm resembles an object or `agent' making a short random-walk. This is very similar to the Growing Network with Redirection model discussed in Section~\ref{sec:Redner} except that the model here employs undirected links which necessitates a random (as opposed to deterministic) walk. Clearly, this algorithm generates trees, some examples of which are illustrated in Figure~\ref{fig:exNN}.

\begin{figure}
\begin{center}
\includegraphics[width=0.95\textwidth]{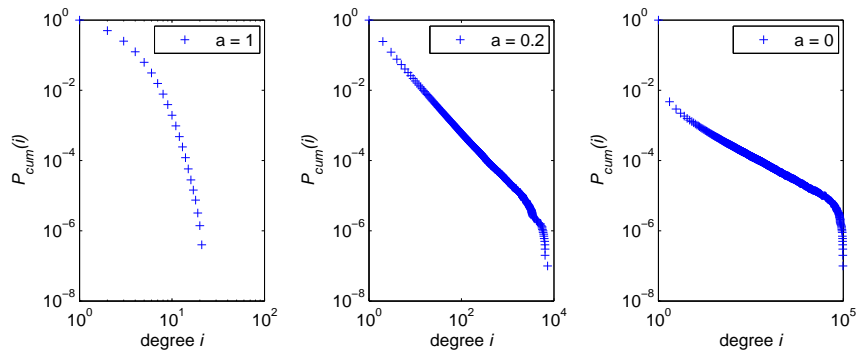}
\caption{ \label{fig:results} The (cumulative) degree distributions for the evolving networks grown to $10^5$ nodes, ensemble averaged over  
100 networks per parameter value. These have been initiated from a 2 node, 1 link component.}
\end{center} 
\end{figure}

The resulting distributions  of vertex degree from implementing this growth algorithm are shown in Figure~\ref{fig:results}. These are ensemble averages over 100 networks per parameter value. Interestingly, $a=0$ yields a network that is dominated by hubs while $a = 1 $ yields the random attachment network. Intermediate values of $a$ yield networks which are neither too ordered nor too disordered. For $a\approx 0.2$, the algorithm generates networks whose degree-distribution closely resembles the BA preferential-attachment network.

% and spokes\footnote{ Consider seeding the algorithm with an existing hub and spoke graph, with our parameter $a = 0$. The  
%probability that a new node will be linked to the hub tends to 1 as the network grows. Indeed, the distribution of vertex  
%degree will approach that of a hub and spoke graph as the node number gets large. This type of network might be considered the  
%most ordered in that the mean free path is minimised (tends to 2).}

Our analysis of the algorithm starts by establishing the attachment probability $\Theta_i$, which in turn requires properly resolving the one-step random walk. The link-space formalism provides us an expression for the probability $P'_i$ associated with performing a  
random walk of length one and arriving at any node of degree $i$. 
\begin{eqnarray}\label{eqn:neighbour}
P'_i&=& P_1 P_{1\rightarrow i}~+~P_2 P_{2\rightarrow i}~+~P_2 P_{2\rightarrow i}~+~P_3 P_{3\rightarrow i}~+~\ldots  ,\nonumber\\
{}&=& \frac{X_1(t) L_{1,i}(t)}{N(t) 1 X_1(t)}~+~\frac{X_2(t) L_{2,i}(t)}{N(t) 2  
X_2(t)}~+~\frac{X_3(t) L_{3,i}(t)}{N(t) 3 X_3(t)}~+~\ldots ,\nonumber\\
{} &=& \frac{1}{N(t)}\sum_k \frac{L_{k,i}(t)}{k}.
\end{eqnarray}
We can reconcile this with the alternative expression
\begin{eqnarray}
P'_i&=&  \frac{i X_i(t)}{N(t)}\langle\frac{1}{k_{nn}}\rangle_i,
\end{eqnarray}
where the average is performed over the neighbours of nodes of degree $i$. With proper normalisation,
\begin{eqnarray}
\langle\frac{1}{k_{nn}}\rangle_i&=& \sum_{k} \frac{L_{i,k}(t)}{k~\sum_{\alpha}L_{i,\alpha}(t)} ,\nonumber\\
{} & =& \frac{1}{i X_i(t)} \sum_{k}\frac{L_{i,k}(t)}{k},
\end{eqnarray}
which clearly yields the same results. Note that this quantity does {\em not} replicate 
preferential attachment, in contrast to what is commonly thought 
~\cite{Saramaki, Vazquez}. We shall perform this process explicitly for the example ($N=11$, $M =10$) network of Figure~\ref{fig:simplenet4} with the link-space matrix
 \begin{eqnarray}
\mathbf{L}&=& \left( \begin{array}{cccc}
 0&1 & 2 &4\\
 1&0 & 0 &1\\
 2&0 & 0 &1\\
 4&1 & 1 &2
 \end{array} \right).
 \end{eqnarray}
Let us consider the probability of arriving at any node of degree $1$ by performing a random walk of length one from a randomly  chosen initial vertex,
\begin{eqnarray}
 P'_1  &=& P_1 P_{1\rightarrow 1}~+~P_2 P_{2\rightarrow 1}~+~P_2 P_{2\rightarrow 1}~+~P_3 P_{3\rightarrow 1}
 ~+~P_4 P_{4\rightarrow 1}, \nonumber\\
 {}&=& 0~+~\frac{1}{11}\frac{1}{2}~+~\frac{1}{11}\frac{2}{3}~+~\frac{1}{11}\Big(\frac{3}{4}+\frac{1}{4}\Big),\nonumber\\
 {}&=& \frac{1}{N}\sum_k \frac{L_{k,1}}{k}.
\end{eqnarray}
This is clearly consistent with the analysis of (\ref{eqn:neighbour}).\\
We can now begin to formulate the master equations for this model. 
Defining $\beta_i$ as
\begin{eqnarray}
\beta_i &=& \frac{1}{i~c_i}\langle \frac{1}{k_{nn}}\rangle_i ,\nonumber\\
{}&=&\frac{\sum_k \frac{L_{i,k}}{k}}{\sum_k L_{i,k}},\nonumber\\
{}&=&\frac{\sum_k \frac{l_{i,k}}{k}}{\sum_k l_{i,k}},
\label{eqn:beta}
\end{eqnarray}
the attachment kernel for the External Nearest Neighbour algorithm becomes
\begin{eqnarray}
\Theta_i&=& a c_i~+~(1-a)\beta_i i c_i.
\label{eqn:xnnkernel}
\end{eqnarray}
Substituting (\ref{eqn:xnnkernel}) into (\ref{eqn:nsmaster}) we obtain for the steady state node degree
\begin{eqnarray}
c_i&=&\frac{c_{i-1}\big(a+(1-a)\beta_{i-1}(i-1)\big)}{1+a+(1-a)\beta_i i}.
\end{eqnarray}
Substituting into (\ref{eqn:linkspacemaster}) we get
\begin{eqnarray}
l_{i,j} &= &\frac{l_{i-1,j}\big(a+(1-a)\beta_{i-1}(i-1)\big)+l_{i,j-1}\big(a+(1-a)\beta_{j-1}(j-1)\big)}
{1+2a+(1-a)(i\beta_i+j\beta_j)}  ~~~~i,j>1 ,\nonumber\\
l_{1,j}&=& \frac{l_{1,j-1}\big(a+(1-a)\beta_{j-1}(j-1)\big)+c_{j-1}\big(a+(1-a)\beta_{j-1}(j-1)\big)}
{1+2a+(1-a)(\beta_1+j\beta_j)}~~~~j>1 ,\nonumber\\
l_{1,1}&=&0.
 \end{eqnarray}
The non-linear terms resulting from $\beta$ mean that a complete analytical solution for $l_{i,j}$ is difficult. We leave this as a future challenge. However, the formalism  can  be implemented in its non-stationary form numerically (iteratively) with reasonable efficiency. As described in Section~\ref{sec:linkspace}, the master equations for the network growth in the link-space are given for ${i,j}>1$ as
\begin{eqnarray}
\langle L_{i,j}(t+1) \rangle & =& L_{i,j}(t)  
+\frac{\Theta_{i-1}(t)L_{i-1,j}(t)}{X_{i-1}(t)}+\frac{\Theta_{j-1}(t)L_{i,j-1}(t)}{X_{j-1}(t)}\nonumber\\
{}&{}&-~\frac{\Theta_{i}(t)L_{i,j}(t)}{X_{i}(t)}-
\frac{\Theta_{j}(t)L_{i,j}(t)}{X_{j}(t)},
\label{eqn:Lspace3}
\end{eqnarray}
and for $i=1,j>1$ as
\begin{eqnarray}
\langle L_{1,j} (t+1)\rangle & =& L_{1,j}(t) + \Theta_{j-1}  +\frac{\Theta_{j-1}(t)L_{1,j-1}(t)}{X_{j-1}(t)}\nonumber\\
{}&{} &- \frac{\Theta_{1}(t)L_{1,j}(t)}{ X_{1}(t)}-
\frac{\Theta_{j}(t)L_{1,j}(t)}{ X_{j}(t)}.
\label{eqn:Lspace4}
\end{eqnarray}
Computationally, it is useful to  rewrite these equations (\ref{eqn:Lspace3}) and (\ref{eqn:Lspace4}) in terms of matrix operations. We define the following matrices and operators:
\begin{eqnarray}
\mathbf{E} &=& \left(\begin{array}{c c c c c}
0&1&0&0&\ldots\\
0&0&1&0&\ldots\\
0&0&0&1&\ldots\\
0&0&0&0&\ddots\\
\vdots & \vdots & \vdots & \vdots& {} \end{array}\right),\nonumber\\
\mathbf{F} &=& \left(\begin{array}{c c c c c}
0&0&0&0&\ldots\\
1&0&0&0&\ldots\\
0&1&0&0&\ldots\\
0&0&1&0&\ldots\\
\vdots & \vdots & \vdots & \ddots& {} \end{array}\right),\nonumber\\
\mathbf{I} &=& \textrm{the identity matrix},\nonumber\\
\mathbf{K} &=& \left(\begin{array}{c c c c c}
1&0&0&0&\ldots\\
1&0&0&0&\ldots\\
1&0&0&0&\ldots\\
1&0&0&0&\ldots\\
\vdots & \vdots & \vdots & \vdots& {} \end{array}\right),\nonumber\\
\end{eqnarray}
\begin{eqnarray}
\otimes &=& \textrm{Hadamard elementwise multiplication such that}\nonumber\\
{}&{}&\textrm{ for $\underline{e}=\underline{f}\otimes \underline{g}$, $e_i = f_ig_i$},\nonumber\\
\oslash &=& \textrm{Elementwise  divide such that for $\underline{e}=\underline{f}\oslash \underline{g}$} ,\nonumber\\
{}&{}& e_i ~= ~\left\{\begin{array}{l r}
f_i/g_i & \textrm{ for $g_i\ne 0$},\\
0 &\textrm{otherwise},\end{array}\right.\nonumber\\
\underline{\gamma}(t)&=&\underline{\Theta}(t)\oslash\underline{X}(t),\nonumber\\
\underline{\rho} &=& \left(\begin{array}{c}
1\\
2\\
3\\
\vdots\end{array}\right),\nonumber\\
\end{eqnarray}
\begin{eqnarray}
\underline{1}&= &\left(\begin{array}{c}
1\\
1\\
1\\
\vdots\end{array}\right),\nonumber\\
\underline{\phi} &=& \underline{1}\oslash\underline{\rho},\nonumber\\
{}&=& \left(\begin{array}{c}
1\\
\frac{1}{2}\\
\frac{1}{3}\\
\vdots\end{array}\right),\nonumber\\
\mathbf{diag(\underline{x})} &=& \left(\begin{array}{c c c c c}
x_1&0&0&0&\ldots\\
0&x_2&0&0&\ldots\\
0&0&x_3&0&\ldots\\
0&0&0&x_4&\ldots\\
\vdots & \vdots & \vdots & \vdots& \ddots \end{array}\right).
\end{eqnarray}
%\end{tiny}

Armed with these useful building blocks, we can represent the link-space master equations, (\ref{eqn:lspace}) as
\begin{eqnarray}\label{eqn:matrix}
\langle \mathbf{L}(t+1)\rangle&=&\Big((\mathbf{F}-\mathbf{I})\mathbf{diag}(\underline{\gamma}(t))+\mathbf{I}\Big)\mathbf{L}(t)+\mathbf{L}(t)\mathbf{diag}(\underline{\gamma}(t))(\mathbf{E}-\mathbf{I})\nonumber\\
{}&{}& +~\mathbf{F}\mathbf{diag}(\underline{\Theta}(t))\mathbf{K} +\mathbf{K}^T\mathbf{diag}(\underline{\Theta}(t))\mathbf{E}.
\end{eqnarray}
Note that we have not lost generality in specifying the attachment kernel $\underline{\Theta}(t) = \underline{\gamma}(t)\otimes\underline{X}(t)$. Although (\ref{eqn:matrix}) looks somewhat intractable, the terms can be easily explained. Premultiplying by $\mathbf{F}$ has the effect of moving all the elements down one place. Postmultiplying by $\mathbf{E}$ moves the elements right. The last two terms represent the $\Theta$ term in (\ref{eqn:Lspace4}).

\begin{figure}
\begin{center}
\includegraphics[width=0.7\textwidth]{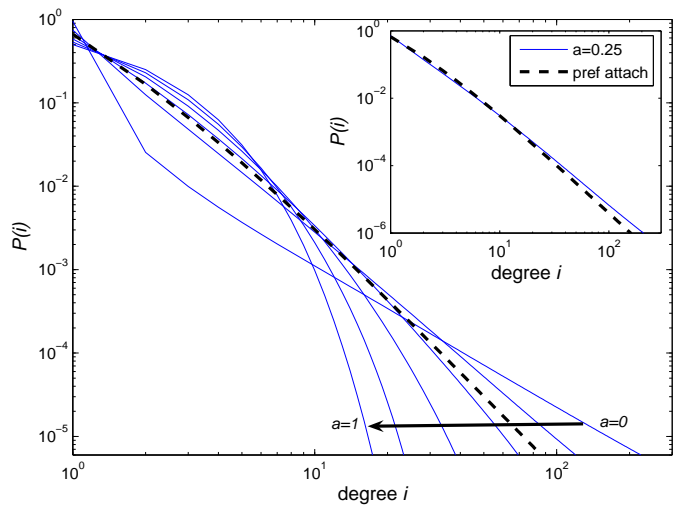}
\caption{ \label{fig:lsresults}Degree distributions generated using the link-space analysis of our one-step algorithm. Dashed line: results for the Barab\'asi-Albert (BA) \cite{Barabasi} preferential-attachment algorithm. Our one-step algorithm ressembles the BA results with parameter value $a=0.25$ (see inset).}
\end{center}
\end{figure}

We can now tailor for the three systems. For random attachment we have
\begin{eqnarray}
\underline{\Theta}(t)&=&\frac{1}{N(t)}\underline{X}(t),\nonumber\\
\underline{\gamma}(t)&=&\frac{1}{N(t)}\underline{1}.
\end{eqnarray}
For preferential attachment we have
\begin{eqnarray}
\underline{\Theta}(t)&=&\frac{1}{2(N(t)-1)}\underline{\rho}\otimes\underline{X}(t),\nonumber\\
\underline{\gamma}(t)&=&\frac{1}{2(N(t)-1)}\underline{\rho}.
\end{eqnarray}
Things start to get a little trickier for the external nearest neighbour algorithm yielding
\begin{eqnarray}
\underline{\Theta}(t)&=&\frac{a}{N(t)}\underline{X}(t) + \frac{(1-a)}{N(t)}\underline{X}(t)\otimes \underline{\rho}\otimes\underline{\beta}(t),\nonumber\\
\underline{\gamma}(t)&=&\frac{a}{N(t)} + \frac{(1-a)}{N(t)} \underline{\rho}\otimes\underline{\beta}(t).
\end{eqnarray}
Recall from (\ref{eqn:beta}) that
\begin{eqnarray}
\beta_i &=&\frac{\sum_k \frac{L_{i,k}(t)}{k}}{\sum_k L_{i,k}(t)},\nonumber\\
\underline{\beta}(t)&=&\big(\mathbf{L}(t)\underline{\phi}\big)\oslash\big(\mathbf{L}(t)\underline{1}\big).
\label{eqn:beta2}
\end{eqnarray}
Actual computation is somewhat more straight-forward than the mathematics suggests and is based upon the assumption that we can interchange $\langle \mathbf{L}(t+1)\rangle$ with $\mathbf{L} (t+1)$ to enable a recursive process for the evolution of the link-space matrix. Whilst this would not be strictly apppropriate to model the evolution of a single realisation of the algorithm or even an ensemble average (each realisation would have a path dependent evolution), if a steady state is reached (albeit approximately) then this will be a steady state of the algorithm. This is a similar assumption to that made in the node-space master equation approach to non-equilibrium networks in Section~\ref{sec:linkspace}.
We seed with $L_{1,1}(1)~ = ~2$ reflecting two connected nodes (such that $N(t) = t+1$ and total links, $M=t$). To account for the finite size of the matrices involved, the link-space is normalised to $2~(N(t)-1)$ at each iteration although in practice, the very edges of the matrix where numbers might fall off have very small values.
We retrieve the degree distribution $\underline{c}$ from the link-space matrix,
\begin{eqnarray}\label{eqn:ci2}
X_i(t) &=& \frac{\sum_k L_{i,k}(t)}{i},\nonumber\\
\underline{X}(t)&=&  \underline{\phi} \otimes \mathbf{L}(t)\underline{1},  \nonumber\\
\underline{c}(t)&=&\frac{1}{N(t)}\underline{X}(t).
\end{eqnarray}
Comparison of the matrices $\mathbf{L}(t+1)$ and $\mathbf{L}(t)$ or the vectors $\underline{c}(t+1)$ and $\underline{c}(t)$ gives an idea of the proximity to the steady state. The results can be seen in Figure~\ref{fig:lsresults} and appear (qualitatively) comparable to the  
numerical numerical simulations of Figure~\ref{fig:results}.% The link-space for the $a=0$ case is illustrated in Figure~\ref{fig:XNNlogloglog}.

\begin{figure}
\begin{center}
\includegraphics[width=0.7\textwidth]{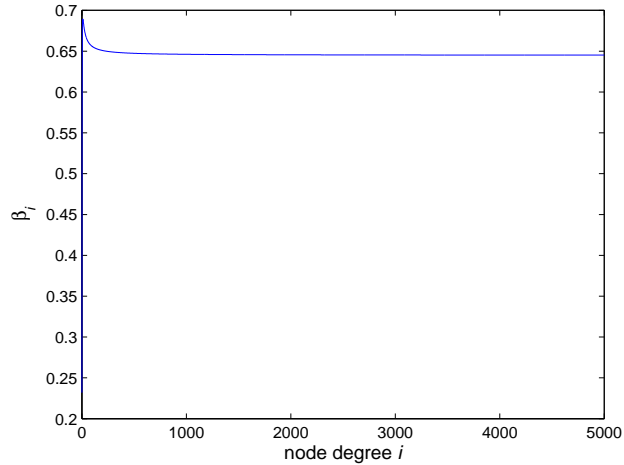}
\caption{ \label{fig:betasf} The value of $\beta_i$ for the preferential attachment algorithm.}
\end{center}
\end{figure}

We can now use the link-space formalism to deduce the parameter value at which our algorithm approximates the BA degree distribution. At this value $a=a_c$, the node-degree distribution goes from higher to lower power-law exponent in relation to the BA model. For this parameter value, the attachment probability to nodes of various degrees is approximately equal for both our mixture algorithm and the BA model. 
Recall from Section~\ref{subsec:BA} and (\ref{eqn:xnnkernel})
\begin{eqnarray}\label{eqn:criticalc}
\frac{i}{2}&\approx& a_c~+~(1-a_c)\beta_i i .
\end{eqnarray}
For large $i$ we have
\begin{eqnarray}
a_c& \approx &1-\frac{1}{2\beta_i}.
\label{eqn:a}
\end{eqnarray}
We can use the exact solution of the link-space for the preferential attachment algorithm (or read off the graph in Figure~\ref{fig:betasf}) to infer $\beta_i$ in the high $i$ limit. Using the first two terms of (\ref{eqn:sfexact}) we obtain $\beta\sim0.66$ (the graph asymptotes to $0.65$) and substituting into (\ref{eqn:a}) gives $a_c\approx0.25$. It is important to note that $\beta_i$ is a function of $i$ (as is evident in Figure~\ref{fig:betasf}) which in turn is non-trivially dependent on $a_c$. Consequently, it is unlikely for (\ref{eqn:criticalc}) to hold for all values of $i$. Whilst the mixture model approximates the preferential attachment distribution at this value, the actual distributions are not identical.

%\newpage

%\newpage
\section{Concluding remarks and discussion}\label{sec:lsconclusion}
In conclusion, we have developed a new formalism which accurately accounts for the node-node linkage correlations in networks. We have employed the formalism to produce analytic solutions to the link-space correlation matrix for the random attachment and BA models of network growth as well as the classical random graph. We have also derived the form of a perfectly non-assortative network for arbitrary degree distribution and illustrated the possiblility of a steady-state degree distribution and link-space for decaying networks. 

We have used the formalism to dispel the myth that a one-step random walk from random initial position in an arbitrary network is the same as selecting a node with probability proportional to its degree. By definition, only a perfectly non-assortative network, as derived in  Section~\ref{subsection:nonassort}, would satisfy this condition. 

The link-space formalism has allowed us to  describe a simple one-parameter network growth algorithm which is able to reproduce a wide variety of degree distributions without any global information about node degrees.  There are clearly many alternative yet similar algorithms which could be designed which might also be analysed with the link-space formalism. These might provide insight into certain real world networks which not only exhibit similar scaling behaviour but might also have similar constraints in their microscopic growth rules.

\end{document}